\documentclass[twocolumn,showpacs,preprintnumbers,amsmath,amssymb]{revtex4}

\usepackage{graphicx}
\usepackage{epsfig}

\usepackage{dcolumn}

\usepackage{longtable}
\usepackage{longtable}

\begin{document}

\title{Static fission properties of actinide nuclei}

\author{P.~Jachimowicz$^{1}$}
\author{M.~Kowal$^{2}$} \email{michal.kowal@ncbj.gov.pl}
\author{J. Skalski$^{2}$}

\affiliation{$^1$ Institute of Physics, University of Zielona G\'{o}ra, Z. Szafrana 4a, 65-516 Zielona G\'{o}ra, Poland}
\affiliation{$^2$ National Centre for Nuclear Research, Pasteura 7, 02-093 Warsaw, Poland}

\begin{abstract}
 Fission barriers heights and excitation energies of superdeformed isomeric
 minima are calculated within the microscopic - macroscopic Woods - Saxon
 model for 75 actinide nuclei for which the experimental data are known.
 State - of - the  - art methods were used: minimization over many deformation
 parameters for minima and the imaginary water flow on many - deformation
 energy grid for saddles, including nonaxial and reflection-asymmetric shapes.
 We obtain 0.82 - 0.94 MeV rms deviation between the calculated and
 experimental barriers and 0.53 MeV rms error in the excitation of superdeformed minima (SD).
  Experimental vs theory discrepancies seem to be
  of various nature and not easy to eliminate, especially if one cares for
 more than one or two observables. As an example, we show that by
 strengthening pairing in odd systems one can partially improve agreement in
 barriers, while spoiling it for masses. We also discuss the "thorium
 anomaly" and suggest its possible relation to a different way
 in which the Ac and Th barriers are derived from experimental data.
\end{abstract}


\maketitle


\section{Introduction}

 Spontaneous fission is one of the two main causes limiting the existence
 of superheavy ($Z>103$) nuclei (SHN).
 Known spontaneous fission half-lives $T_{1/2}^{sf}$ of SHN are mostly in
 the range of ms to seconds for even - even, and  10 s - 1 h for odd ones
 \cite{Hessberger2017}.
 In actinides, $T_{1/2}^{sf}$ show a rapid rise with a decreasing proton
 number, for example: $T_{1/2}^{sf}\approx$ 8 s for $^{252} \rm{No}$,
 ~ 86 y for $^{252} \rm{Cf}$, ~ 10$^{14}$ y for $^{241} \rm {Am}$, and
 reaches
 $\sim 10^{19}$ y for the fissile $^{235} \rm {U}$ \cite{Hessberger2017}
 which may be considered practically stable against spontaneous fission.
  When viewed in the picture of quantum tunneling, such enormous
 differences result from  2 - 3 MeV differences in energy landscapes.
 Therefore, evaluation of fission rates requires a rather precise description
 of the potential energy surfaces.
 Till now, still the most effective way to calculate the latter are the
 semi-phenomenological microscopic-macroscopic methods in which the smooth
 (macroscopic) part of the energy and single - particle potential are
 separately fitted to, respectively,  the bulk and single - particle nuclear
 data.

 Using such an approach based on a deformed Woods-Saxon single - particle
potential \cite{Cwiok1987} and the  Yukawa-plus-exponential macroscopic energy \cite{Krappe1979}
we have recently \cite{Jachimowicz2017_II} systematically calculated static fission barrier
 heights $B_f$ for 1305 heavy and SH nuclei beyond berkelium, including even-even,
odd-even, even-odd and odd-odd systems. In this paper we report a similar study
 for actinide nuclei for which experimental fission data are available.
 Energy surfaces of actinides show long barriers and various saddles and their
  determination requires accounting for many collective deformations; this leads to an
 extensive and time - consuming numerical effort.
 The main goal of our present work is to compare calculated
"inner" and "outer" static fission barriers with empirical estimates.
 Having many - dimensional energy landscapes, without much effort we
 can also find the location and excitation of super-deformed minima.

 Fission barriers heights are model-dependent quantities, and at the same time,
  very useful theoretical constructs related to fission data. It has to be
 remembered that while these data are usually obtained from the neutron -
 induced fission reactions which involve a few MeV excitation of the
 fissioning system, theoretical calculations are performed mostly for
 adiabatic configurations. Still, it is interesting to
 compare empirical and evaluated static fission properties and
 try to understand them.
 The present work is an extension of our previous studies on barriers
 in even-even actinides
 \cite{Kowal2010,Jachimowicz2012} to
 odd-$A$ and odd-odd nuclei, while using enlarged spaces of collective deformations.
 Of particular importance is the search for saddle points by using the
 immersion water flow technique (in the study of fission barriers first used in
 \cite{Mamdouh1998}), which, in principle, should
 save us from inaccuracies of the minimization method, as explained in
 \cite{Myers96,Moller2000,Moller2009}.
 From the present analysis one can reckon the quality
  of our micro-macro approach and form some idea about its predictive
  power in the region of SHN.

 Systematic calculations including odd-$A$ and odd-odd actinides, with
 inner and outer barriers, from actinium to californium, are rather
 scarce. We are aware of results of
 P. M\"oller and coworkers in \cite{Moller2009} and those
 of the HFB14 model by S. Goriely et al. \cite{Goriely2007}. On the
 other hand, there are many published calculations of fission barriers
 in even-even actinides, performed within various micro-macro (e.g the recent work \cite{POM2018}), and
 mean-field models. Some of the latter studies contain a
 careful analysis of various approximations and/or corrections,
 like, for example \cite{BONNEAU04} based on the SkM* force,
 \cite{Delaroche2006,Robledo2014,Lemaitre2018} using Gogny models,
 and \cite{Abusara2010,Zhou2014} using relativistic functionals.
 They widely differ in applied methodology as to the saddle point
 determination and included energy corrections. Sometimes they involve
 arbitrary prescriptions, like for example, for the so-called collective
 energy corrections. Some of these results are astonishingly (taking
  into account applied approximations) close to the experimental estimates.

  We would like to stress that experimental data on fission barriers
  or isomers were never used to fit the parameters of our model.
  Likewise, no single adjustment was made in the present work
  to improve the agreement of calculated fission barrier heights and
  excitation energies of second minima with their experimental estimates.
   Examples of, and comments on changes in calculated quantities
 introduced by modifications of selected parameters of the model are
 presented only to provide some orientation on their interrelation.

\section{Method of the calculation}

We used exactly the same microscopic-macroscopic approach as in our previous global
calculations of static fission barrier heights for the heaviest nuclei $98 \leqslant Z \leqslant 126$,
\cite{Jachimowicz2017_II}. Thus, the Yukawa-plus-exponential model was taken for the macroscopic part of the energy,
and the Strutinsky shell correction, based on the Woods-Saxon single-particle potential was used
for its microscopic part. All parameters used in the present work, that have been
fixed previously (see \cite{Jachimowicz2017_II} and references therein), were kept unchanged.
Also the pairing correlations were taken into account in our model within
the standard BCS approach without any specially adjusted parameters. Additionally, for systems with odd numbers
of protons, neutrons, or both, we used a standard blocking method. Other details of the approach
are also specified in \cite{Jachimowicz2017_II, Jachimowicz2014}.

To describe nuclear shapes we used a standard $\beta$ parametrization which consists in the
expansion of the nuclear radius vector in spherical harmonics:
\begin{eqnarray} \label{parametryzacjaB}
R(\vartheta ,\varphi)= c R_0\{
1+\sum_{\lambda=1}^{\infty} \sum_{\mu=-\lambda}^{+\lambda} \beta_{\lambda \mu} {\rm Y}_{\lambda \mu} (\vartheta ,\varphi) \},
\end{eqnarray}
\\
where $c$ is the volume-fixing factor depending on deformation and $R_0$ is
the radius of a spherical nucleus. For large elongations this parametrization
cannot be very efficient, however our tests and comparisons with other
 parameterizations, as the modified Funny-Hills \cite{Pomorski2006} or three
 quadratic surfaces, e.g. \cite{Moller2009}, indicate that it is still good
 and effective in the region of the second barrier.
 One should realize that the superdeformed and the second saddle shapes
 in actinides are still rather compact.
 On the other hand, with an increasing elongation the relative importance of
 different spherical harmonics changes along the fission path.
 For this reason our searches for minima and the first and second saddles were
 performed independently, by using different deformation sets. This allowed
 to reduce the computational effort, making calculations feasible while still
 preserving the reliability of the results.
 The shape parameterizations used in different regions of potential energy
 surfaces (PES) are detailed below.

\subsection{Ground states \& second minima}

 For nuclear ground states (g.s.), based on our previous tests and results \cite{Jachimowicz2017_I},
we confined our analysis to axially-symmetric shapes
with expansion of the nuclear radius (\ref{parametryzacjaB}) truncated at $\beta_{80}$ :
\begin{eqnarray} \label{gs}
R(\vartheta ,\varphi) = c R_0\{
1 & + & \beta_{2 0} {\rm Y}_{2 0} + \beta_{3 0} {\rm Y}_{3 0} + \beta_{4 0} {\rm Y}_{4 0} \nonumber \\
  & + & \beta_{5 0} {\rm Y}_{5 0} + \beta_{6 0} {\rm Y}_{6 0} + \beta_{7 0} {\rm Y}_{7 0} \\
  & + & \beta_{8 0} {\rm Y}_{8 0} \},  \nonumber
\end{eqnarray}
 where here and in the following the angular dependence of spherical harmonics is suppressed.
 Therefore, in this case, the energy was minimized over 7 degrees of freedom
 specified in (\ref{gs}), by using the conjugate gradient method. To avoid
 falling into local minima, the minimization was repeated dozens of times
 for each nucleus, with randomly selected starting deformations.
 For odd systems, the additional minimization over configurations was
 performed at every step of the gradient procedure.

 Exactly the same procedure and deformation space (\ref{gs}) were used to
 determine isomeric, superdeformed (SD) minima and their excitation energies
 $E^{*}$ relative to the ground states.
 Starting points did not have to be guessed as this minimization was
 done after we had calculated the full energy grids (see the point C).
 The gradient method is, however, more accurate and therefore,
 in order to determine the location of these minima as precisely as possible,
 we have applied it for the relevant points read from the energy maps.
 As it turned out, the obtained secondary minima are exclusively
 mass-symmetric - their deformations $\beta_{30}$,
 $\beta_{50}$, $\beta_{70}$ are equal zero.
 In addition, we have also systematically checked that the nonaxiality plays
 no role in the region of SD minima.
 This result is in line with our previous  conclusions in
 \cite{Jachimowicz2012}.

\subsection{First saddle points}

Much more demanding is to find all saddle points on energy grids (hypercubes).
It is well known that in the region of the first barrier the triaxiality is
 very important \cite {CWIOK92,
CWIOK96,GHERGH99,DUTTA00,DECHARGE2003,BONNEAU04,Cwiok05,DOB2007, KOWAL2009}.
So, in order to find proper first saddle points we used a five dimensional
 deformation space, with the expansion of the nuclear radius:
\begin{eqnarray} \label{sp1}
R(\vartheta ,\varphi) = c R_0\{
1 & + & \beta_{2 0} {\rm Y}_{2 0} + {\beta_{2 2} \over {\sqrt{2}}} \lbrack {\rm Y}_{2 2} + {\rm Y}_{2 -2} \rbrack   \\
  & + & \beta_{4 0} {\rm Y}_{4 0} +  \beta_{6 0} {\rm Y}_{6 0} +
 \beta_{8 0} {\rm Y}_{8 0} \} , \nonumber
\end{eqnarray}
where the quadrupole non-axiality $\beta_{2 2}$ is included explicitly.
For each nucleus we generated the following 5D grid of deformations:
\begin{eqnarray} \label{sp1grid}
\beta_{2 0} & = & \phantom {-} 0.00 \; (0.05)   \; 0.60 \nonumber \\
\beta_{2 2} & = & \phantom {-} 0.00 \; (0.05)  \; 0.45 \nonumber \\
\beta_{4 0} & = &           -  0.20 \; (0.05)  \; 0.20 \\
\beta_{6 0} & = &           -  0.10 \; (0.05)  \; 0.10 \nonumber \\
\beta_{8 0} & = &           -  0.10 \; (0.05)  \; 0.10 \nonumber
\end{eqnarray}
of $29\:250$ points (nuclear shapes);
 the numbers in the parentheses specify the grid steps. Additionally, for
 odd - and odd - odd nuclei, at each grid point we were looking for low-lying
 configurations by blocking particles on levels from the 10-th below to the
 10-th above the Fermi level. Then, our primary grid (\ref{sp1grid})
 was extended by the fivefold interpolation in all directions. Finally,
 we obtained the interpolated energy grid of more than 50 million points.
 To find the first saddles on such a giant gird we used the imaginary water
  flow method, (see e.g. \cite{Moller2009,Jachimowicz2017_II}).
 It is worth mentioning that for all those saddles we carried out an additional
 test of their stability against mass-asymmetry.
 This was done by a 3 dimensional energy minimization with respect to:
 $\beta_{30},\beta_{50}$ and $\beta_{70}$,
 around each saddle. The result of this minimization
 indicates no effect of the mass asymmetry at the first saddle point, similarly
 as in our previous study of superheavy nuclei \cite{Jachimowicz2017_II}.
 This justifies the omission of the mass-asymmetric shapes
 in the definition (\ref{sp1}) of the nuclear radius.

\subsection{Second saddle points}

To determine the second saddle point, we used the following expansion of the nuclear radius:
\begin{eqnarray} \label{sp2}
R(\vartheta ,\varphi) = c R_0\{
1 & + & \beta_{2 0} {\rm Y}_{2 0} + \beta_{3 0} {\rm Y}_{3 0} + \beta_{4 0} {\rm Y}_{4 0} \nonumber \\
  & + & \beta_{5 0} {\rm Y}_{5 0} + \beta_{6 0} {\rm Y}_{6 0} + \beta_{7 0} {\rm Y}_{7 0} \\
  & + & \beta_{8 0} {\rm Y}_{8 0} \},  \nonumber
\end{eqnarray}
 where additionally the dipole distortion $\beta_{1 0}$, important for large
 elongations with a sizable mass asymmetry \cite{Kowal2012,Jachimowicz2013},
 has been used. It was treated as a constraint: for each set of other
 deformations the value of $\beta_{1 0}$ was fixed by setting the
 center of mass of the nucleus to zero (the origin of coordinates).
 The imaginary water flow analysis was performed on the 7-dimensional grid.
The following values of deformation parameters have been used on the grid:
\begin{eqnarray} \label{sp2grid}
\beta_{2 0} & = & \phantom {-} 0.15 \; (0.05)  \; 1.50 \nonumber \\
\beta_{3 0} & = & \phantom {-} 0.00 \; (0.05)  \; 0.45 \nonumber \\
\beta_{4 0} & = &           -  0.15 \; (0.05)  \; 0.35 \nonumber \\
\beta_{5 0} & = &           -  0.20 \; (0.05)  \; 0.25   \\
\beta_{6 0} & = &           -  0.15 \; (0.05)  \; 0.15 \nonumber \\
\beta_{7 0} & = &           -  0.15 \; (0.05)  \; 0.15 \nonumber \\
\beta_{8 0} & = &           -  0.10 \; (0.05)  \; 0.10, \nonumber
\end{eqnarray}
with the steps given in the parentheses. These made a grid of $7\:546\:000$ i
 points for a given nucleus.
  In this case, we could afford a twofold interpolation. However, in performed
 tests we found that it had only a minor effect on heights of the second
 barriers. Therefore, we performed calculations on the original grid.
 As previously, for odd systems, the minimization over configurations (by
 blocking particles on levels from the 10th below to the 10th
above the Fermi level) was performed at each point of the grid (\ref{sp2grid}).
Moreover, in selected nuclei we checked that the quadrupole nonaxiality,
 omitted in (\ref{sp2}), plays a minor role at elongations close to the second
 saddle. The similar conclusion we found in \cite{Jachimowicz2012} for even-even actinides.
 Therefore, our analysis confined here
 to only axially symmetric shapes, should still be reliable.

\section{Results}

\subsection{Ground state masses} \label{subsection:Groundstatemasses}

\begin{figure*}[h]
\includegraphics[scale=0.33]{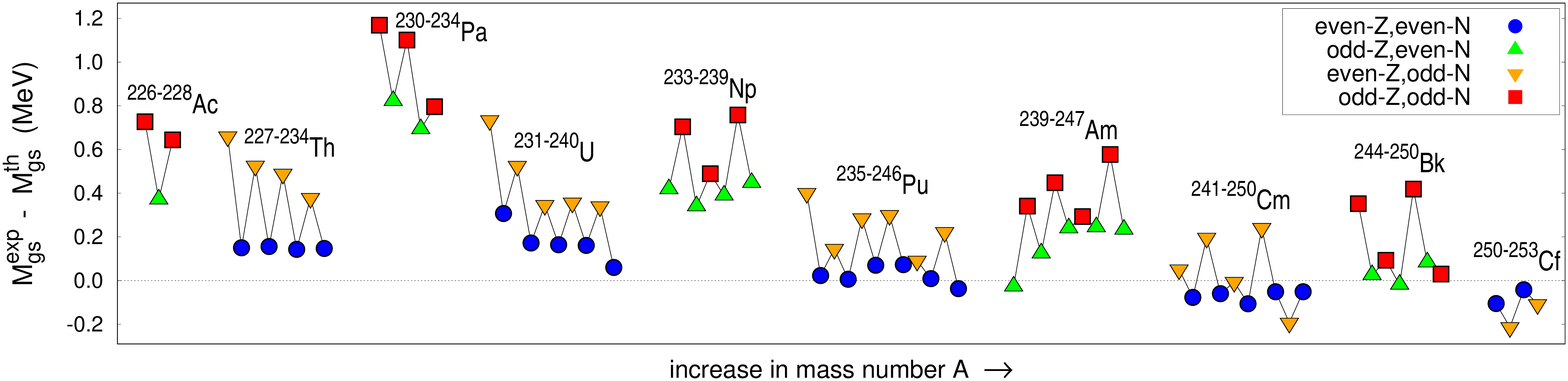}
\caption{
Difference between experimental \cite{Audi2003} and theoretical (our)
ground state masses for considered 75 actinides in separate groups
of even-even, odd-even, even-odd, and odd-odd nuclei.
}
\label{fig:GS}
\end{figure*}

 The present model was used for a description of the experimental g.s. masses
of 252 nuclei with $Z \geqslant 82$ in \cite{Jachimowicz2014}. This
 was an extension to odd-$A$ and odd-odd nuclei of the version used
 previously for even-even heavy nuclei, whose parameters were fixed by a mass
 fit in \cite{Muntian2001}.
 Although excitation energies and fission barriers are calculated
 relative to g.s. energy, it makes sense to see the quality of the mass
 fit.
 Differences between measured \cite{Audi2003} and calculated g.s. masses are
 shown in \mbox{Fig. \ref{fig:GS}}; even-even,
 odd-even, even-odd, and odd-odd nuclei are indicated by different colors and
 shapes.
 A quite pronounced odd-even staggering in these differences signals
 a different degree of accuracy in reproducing g.s. masses in various
 groups of nuclei.
 The differences are the smallest for even-even nuclei (this was the result of
 the original fit in \cite{Muntian2001}), while
 the largest, up to 1.2 MeV, occur for odd-odd systems, especially for
 Pa isotopes.
 One can also notice a systematic underestimate of the
 experimental masses in lighter elements that means that the calculated
 binding energies (meaning their absolute values) are too large there.
 Thus, we have overbinding in lighter elements, which is more pronounced in
 odd and odd-odd nuclei

 One could think that this even-odd difference in the mass fit might be
 related to the blocking method which leads to a too
 strong reduction in the pairing gap. However, one should notice that the
 binding in odd nuclei is overestimated more than in even ones, so it has
 another cause. One can notice that
 an increase in pairing strength for all nuclei would decrease
 the staggering {\it in the binding error} between odd and even ones (as a
 stronger pairing increases the effect of blocking on energy) but would also
 deteriorate the relatively good mass fit for even-even nuclei.

 On the other hand, the blocking effect may cause too high barriers in the odd
 systems, as a weaker pairing produces higher fission barriers.
 To compensate for this one could assume a slightly
 stronger pairing interaction for odd-particle-numbers.
 Such a test will be discussed in subsection III F. Although the effect is not
 negligible, to keep the consistency with our previous papers, all other
 presented results were obtained with the previously used parameters
 (including pairing).


\subsection{Fission isomers}

  Fast fissioning states discovered in actinides by
  Polikanov et al. \cite{Pol} were soon afterwards interpreted as the
  secondary minima at large elongation in corresponding nuclei
 \cite{FD66,St67}. Their existence disclosed a double - hump shape of fission
 barriers.
  The lowest and excited nuclear states at second minimum are extremely
 short-lived, with characteristic half-lives in the range of 10 ps to 10 ms,
 what makes their experimental study very difficult.
 A more detailed structure of these states (quadrupole moments, energy levels)
 is known only in a few nuclei. Recently,
 quite extensive experimental results were collected on many energy levels
 in the second minimum of $^{240}$Pu \cite{240PuIIa,Gassmann,Thirolf}.

 Deformation of shape isomers is a primary information for any description of
 nuclear structure existing there. The excitation energies of these minima
 have effect on calculations of barrier transmission.
 Thus, it is interesting to compare the experimental vs calculated excitation
 energies, $E^{*th}_{II} = E^{th}_{II} - E^{th}_{gs}$.
 It is also a test of the predictive power of our model, as its parameters
 were not adjusted to these data.

\begin{figure*}[h]
\includegraphics[scale=0.33]{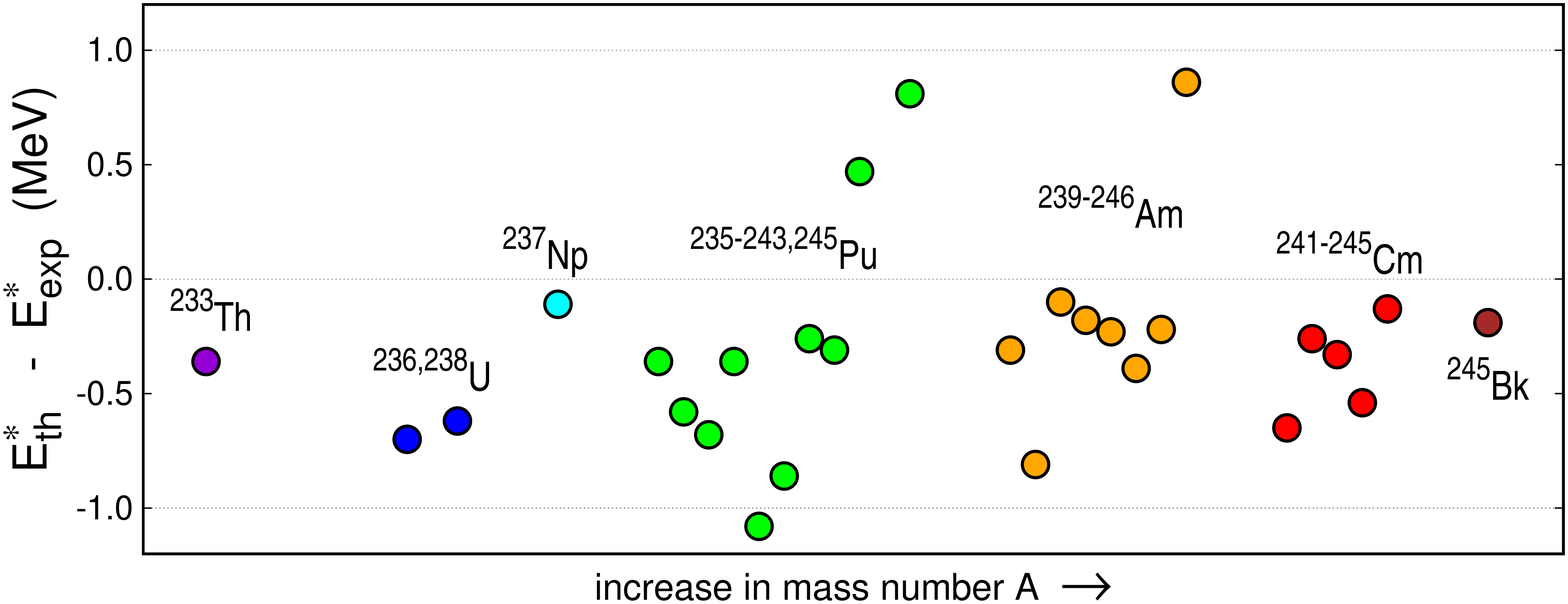}
\caption{Excitation energy of SD minimum; theory vs experiment.}
\label{ESD}
\end{figure*}

 This comparison is provided in Table \ref{TABLE_TOT} and shown in Fig. \ref{ESD}.
 The experimental data \cite{Singh1996} are concentrated in Pu, Am and Cm
 nuclei; only few fission isomers are known in the lighter and in heavier
 actinides. One can also remark that the isomer excitation energies are
 measured with widely varying accuracy, many with uncertainties 0.2 - 0.4 MeV.

  The calculated second minima in most cases lie too low and the spread of
 calculated points around experimental values is quite large.
 The mean deviation of theoretical values from experimental ones is 0.46 MeV while $\delta_{rms} =0.53$ MeV.
 The largest difference of 1.1 MeV between our results and experimental data
  occurs in $^{239} \rm{Pu}$. However, when one
 discards the largest discrepancies: too low $E^{*th}_{II}$ in
 $^{239,240}$Pu and $^{240}$Am, and too high $E^{*th}_{II}$ in
 $^{245}$Pu and $^{246}$Am, the remaining calculated points lie within $\sim$
 0.5 MeV from the experimental ones. If one, additionally, allows for
 experimental uncertainties, the overall agreement looks better.
 Still, it is better than achieved by most of the various Skyrme density
 functionals for which differences between theoretical
 $E^{*th}_{II}$ and experimental $E^{*exp}_{II}$ excitation
 energy of the second minimum can be as high as 4 MeV \cite{Nikolov}.

\begin{figure*}[h]
\includegraphics[scale=0.33]{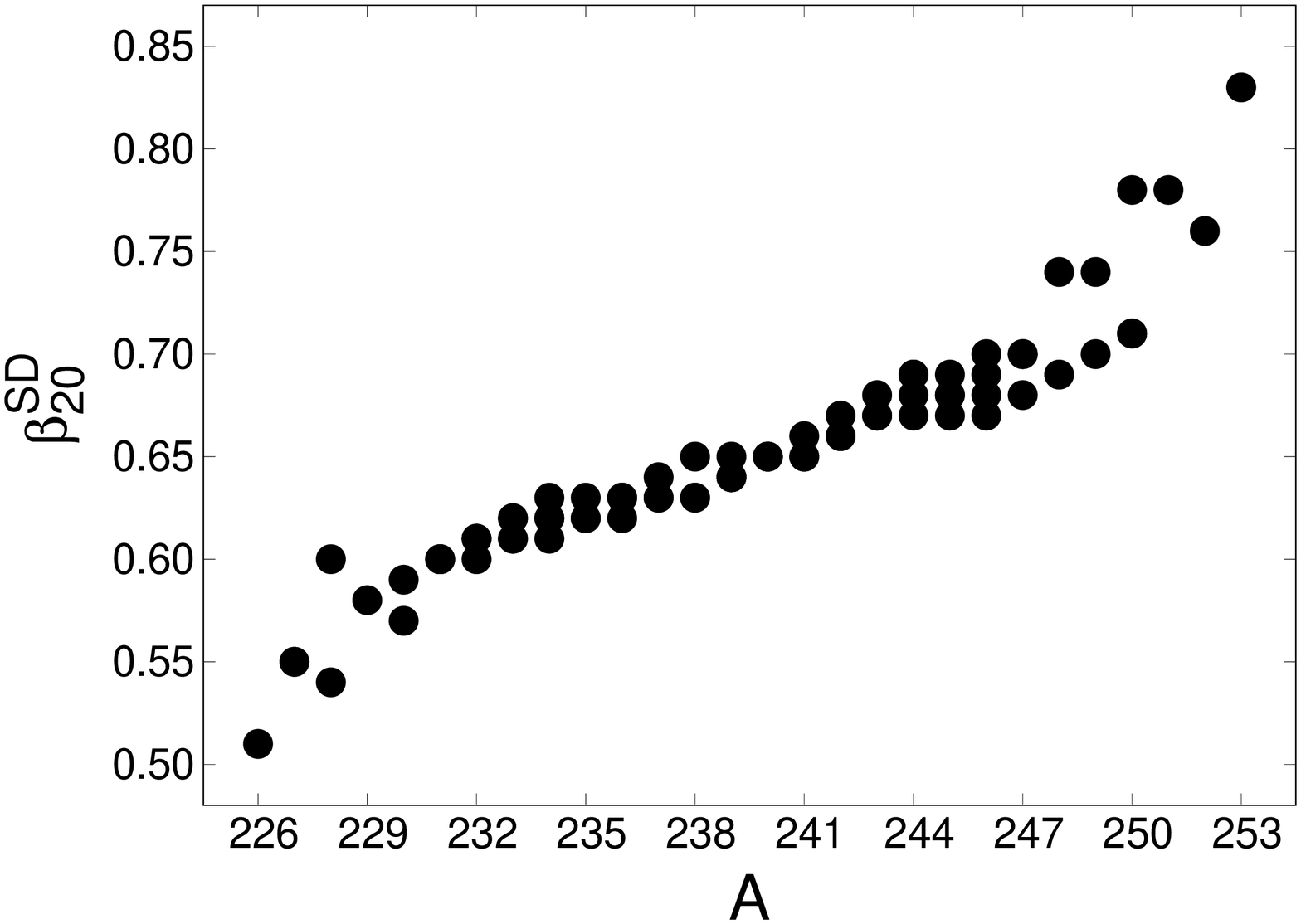}
\caption{Qudrupole deformation for SD minimum.}
\label{b2SD}
\end{figure*}

 Some qualitative features of the data are reproduced by our calculations.
 For example, the obtained excitation of the SD minimum in $^{233}\rm{Th}$
 is smaller than in $^{236,238} \rm{U}$ - as it is in experiment
\cite{Singh1996}. One can also notice that both experimental and theoretical
 $E^{*}_{II}$ are relatively low in the vicinity of $N \approx 147$
 (unfortunately, there is only one data point for Bk - $^{245} \rm{Bk}$).
 Quadrupole deformations $\beta^{SD}_{20}$ of SD minima are shown in
 Fig. \ref{b2SD}. One can see that this variable changes
 linearly with $A$.
 Such a behavior of $\beta^{SD}_{20}$, together with a more steady position,
 $\beta_{20}\approx$ 0.75 - 0.85, of the second saddle, is partially
 responsible for a reduction of the outer barrier width with increasing $A$.
 Although the effect seems small, it can significantly influences
 tunnelling probabilities, i.e. fission half-lives.

\subsection{First fission barrier heights}

\begin{figure*}[h]
\includegraphics[scale=0.4]{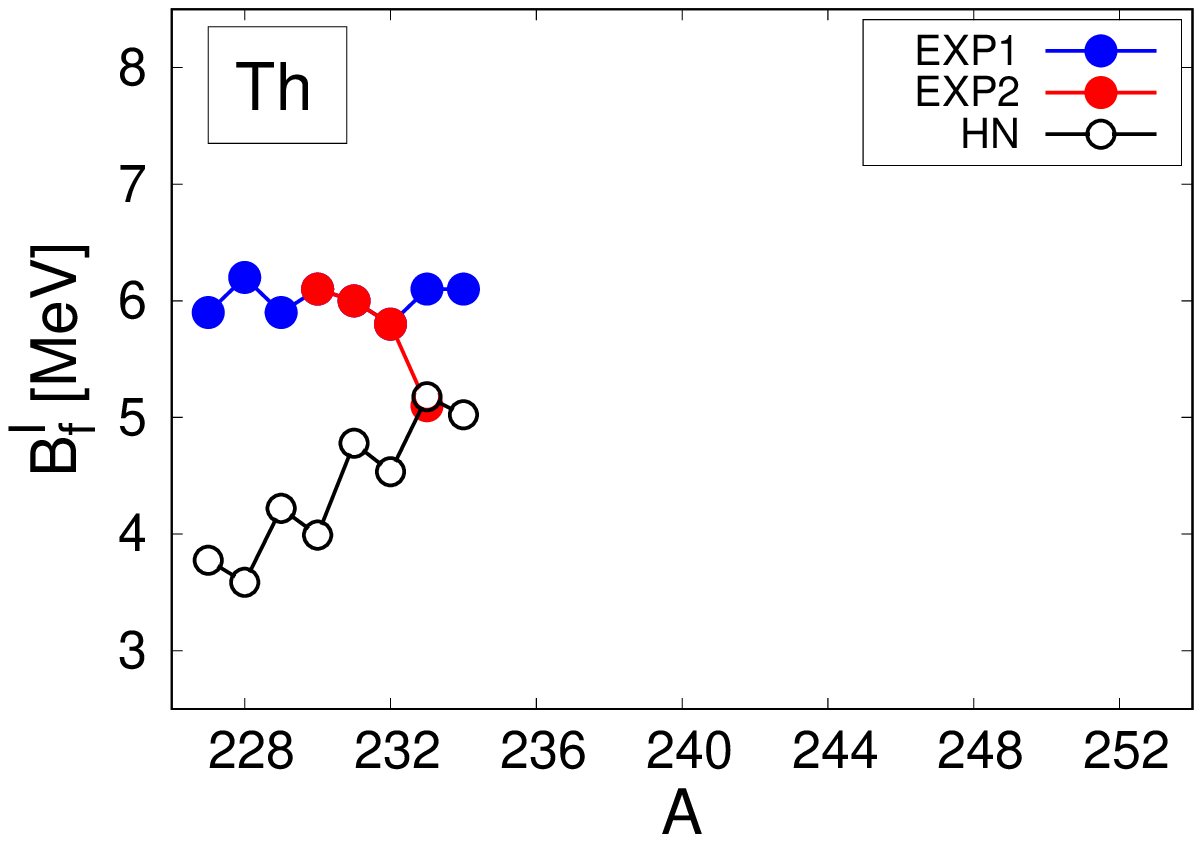}
\includegraphics[scale=0.4]{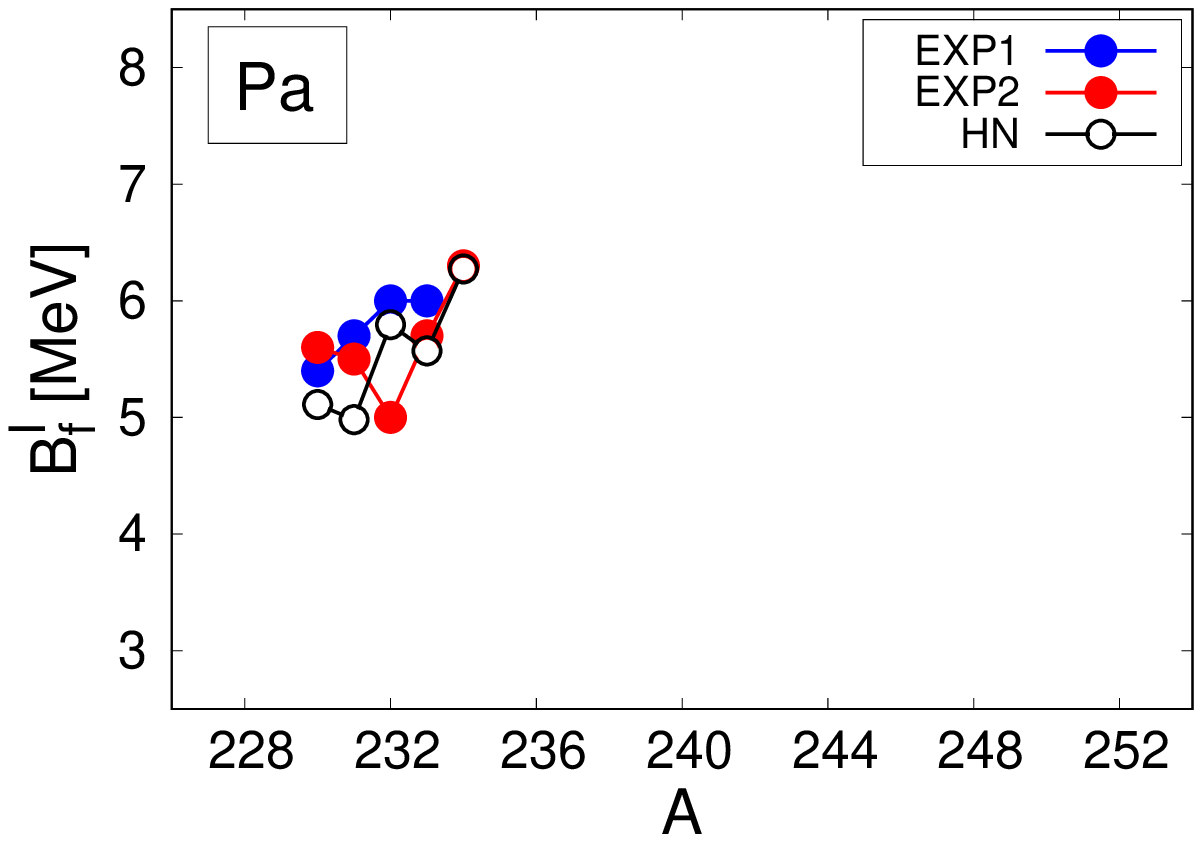}
\includegraphics[scale=0.4]{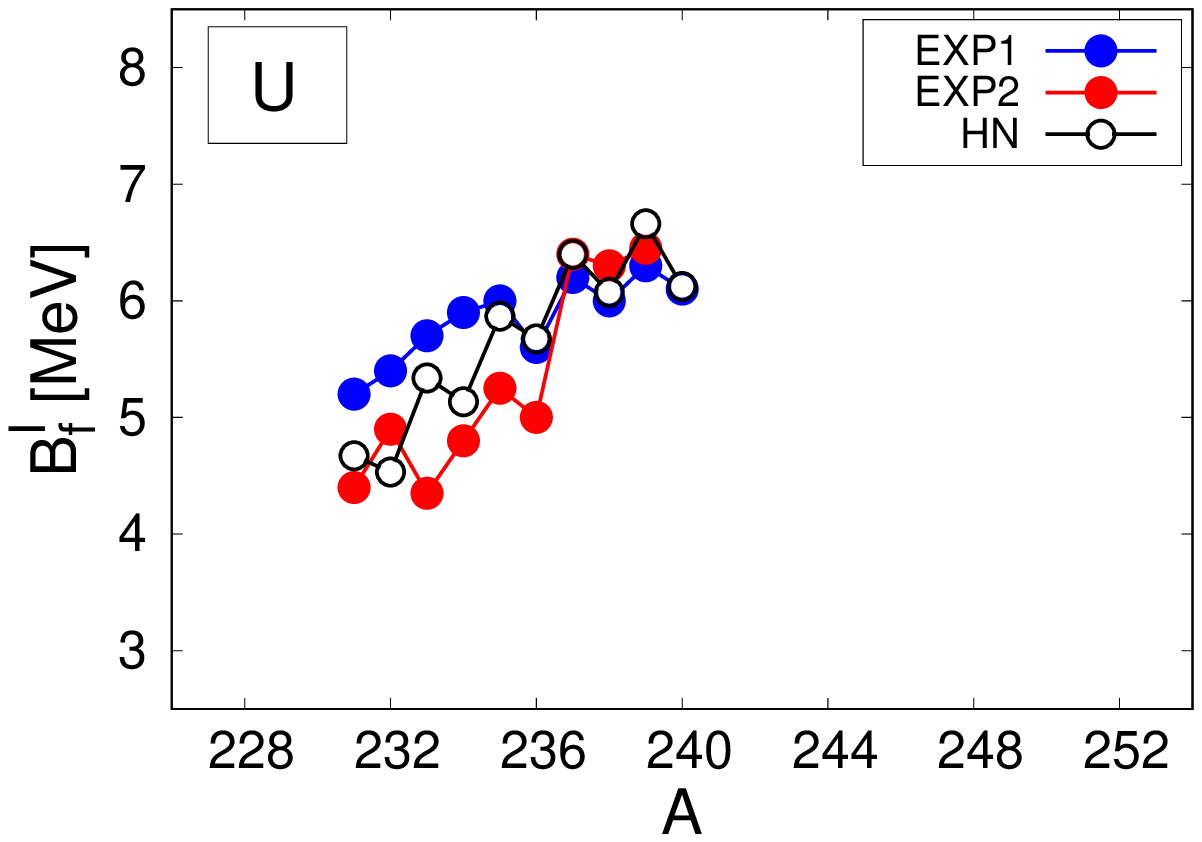}
\includegraphics[scale=0.4]{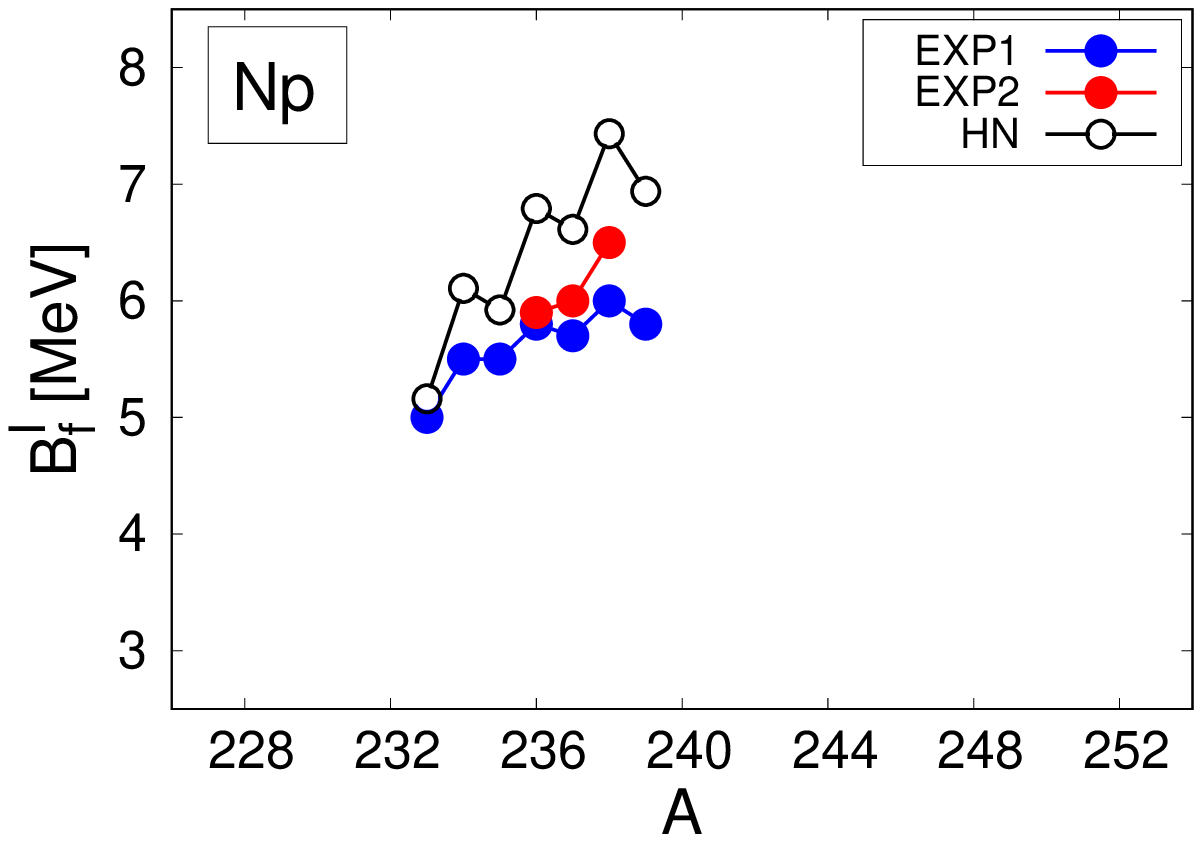}
\includegraphics[scale=0.4]{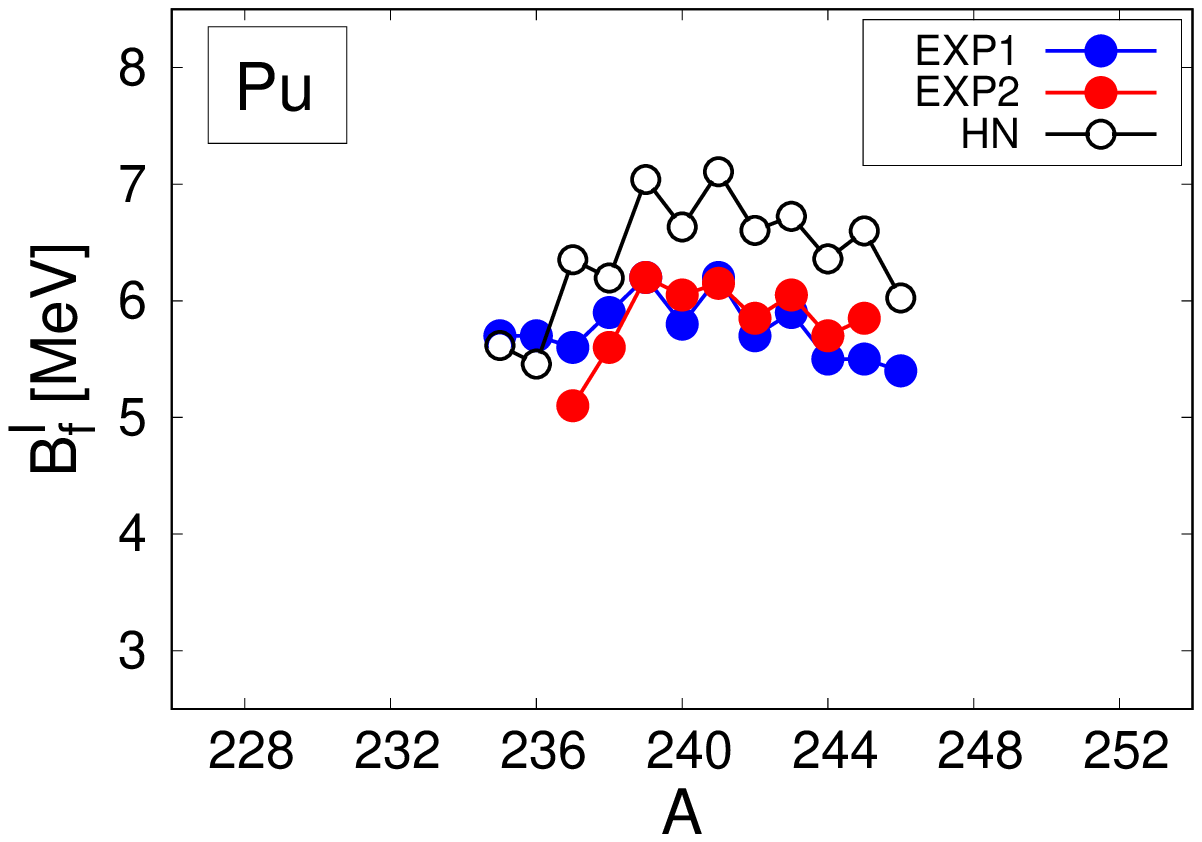}
\includegraphics[scale=0.4]{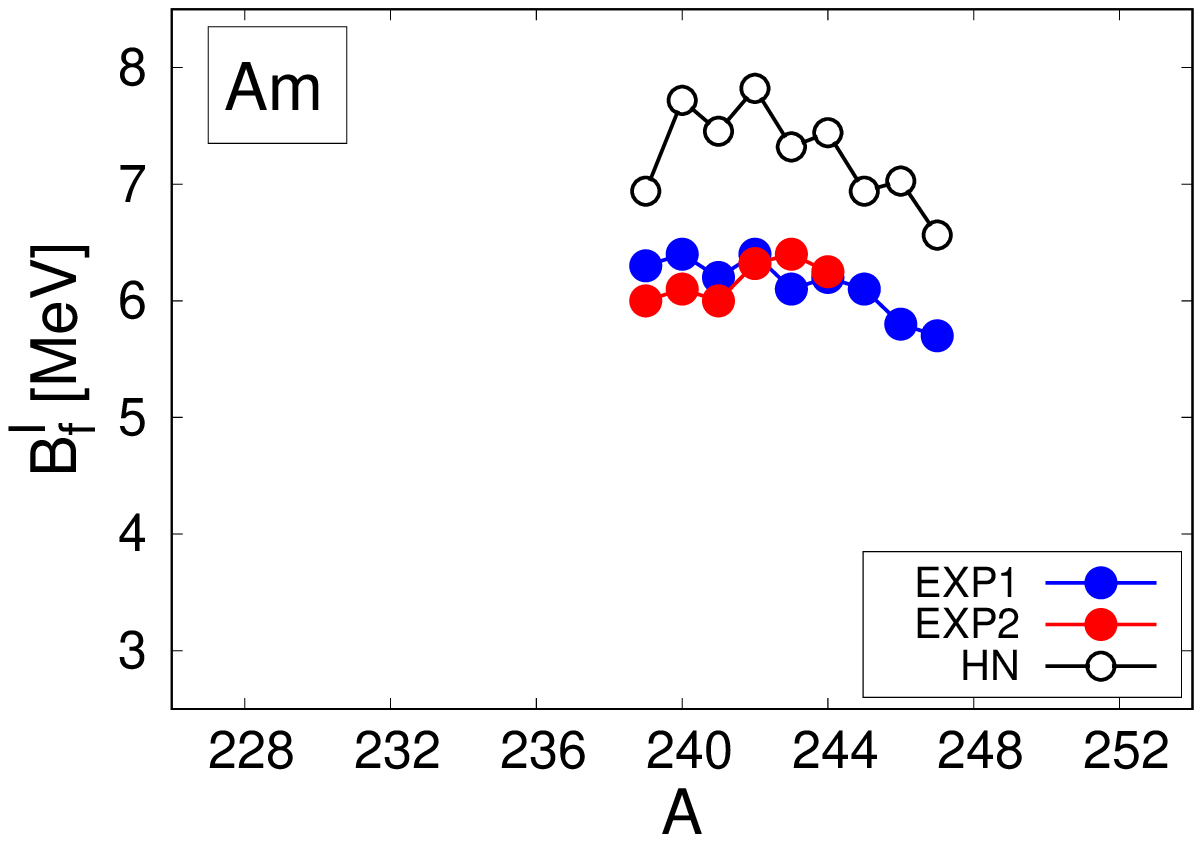}
\includegraphics[scale=0.4]{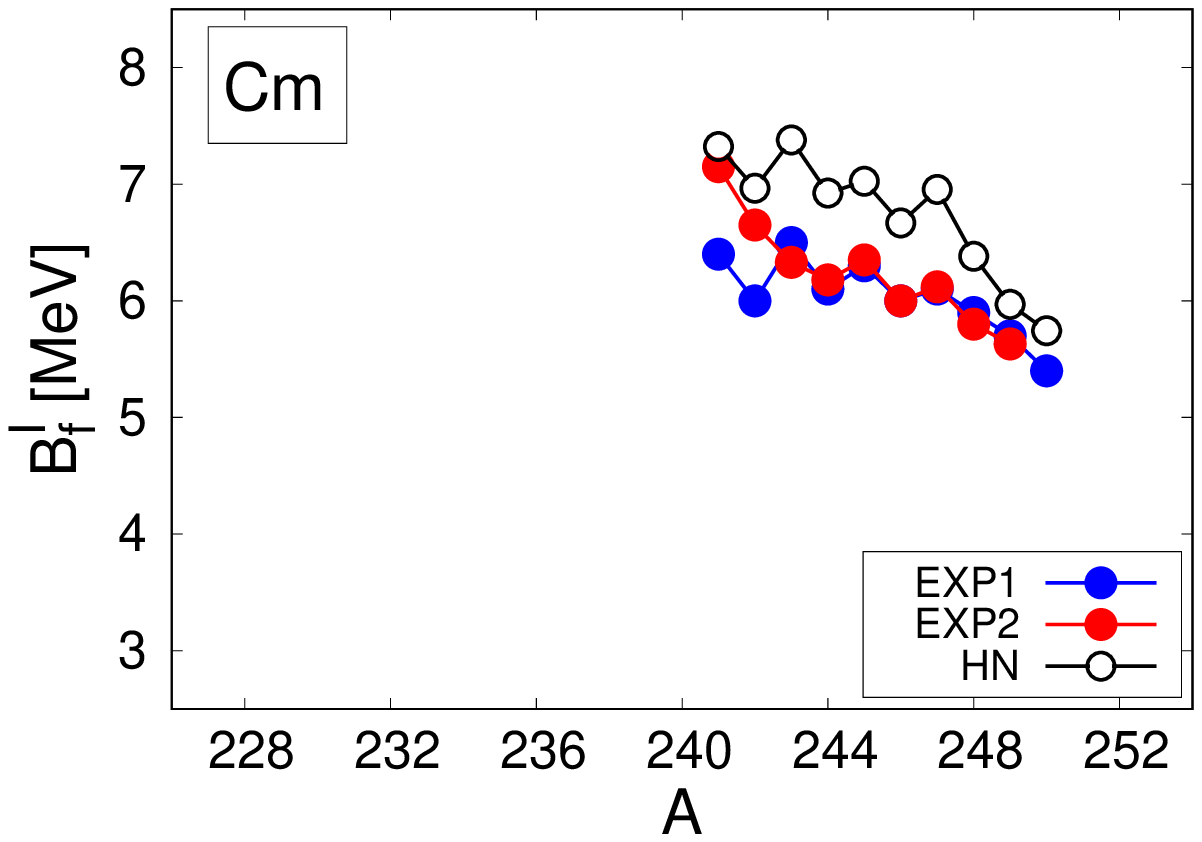}
\includegraphics[scale=0.4]{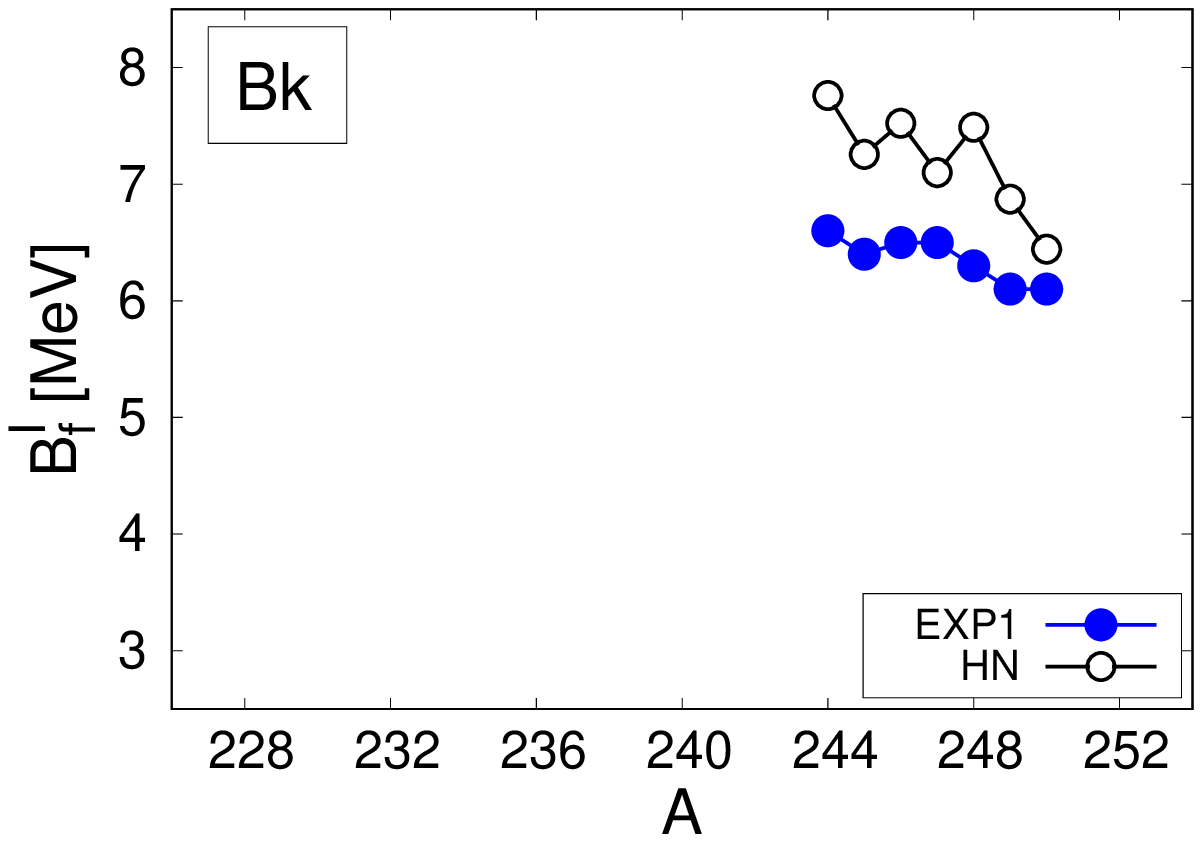}
\includegraphics[scale=0.4]{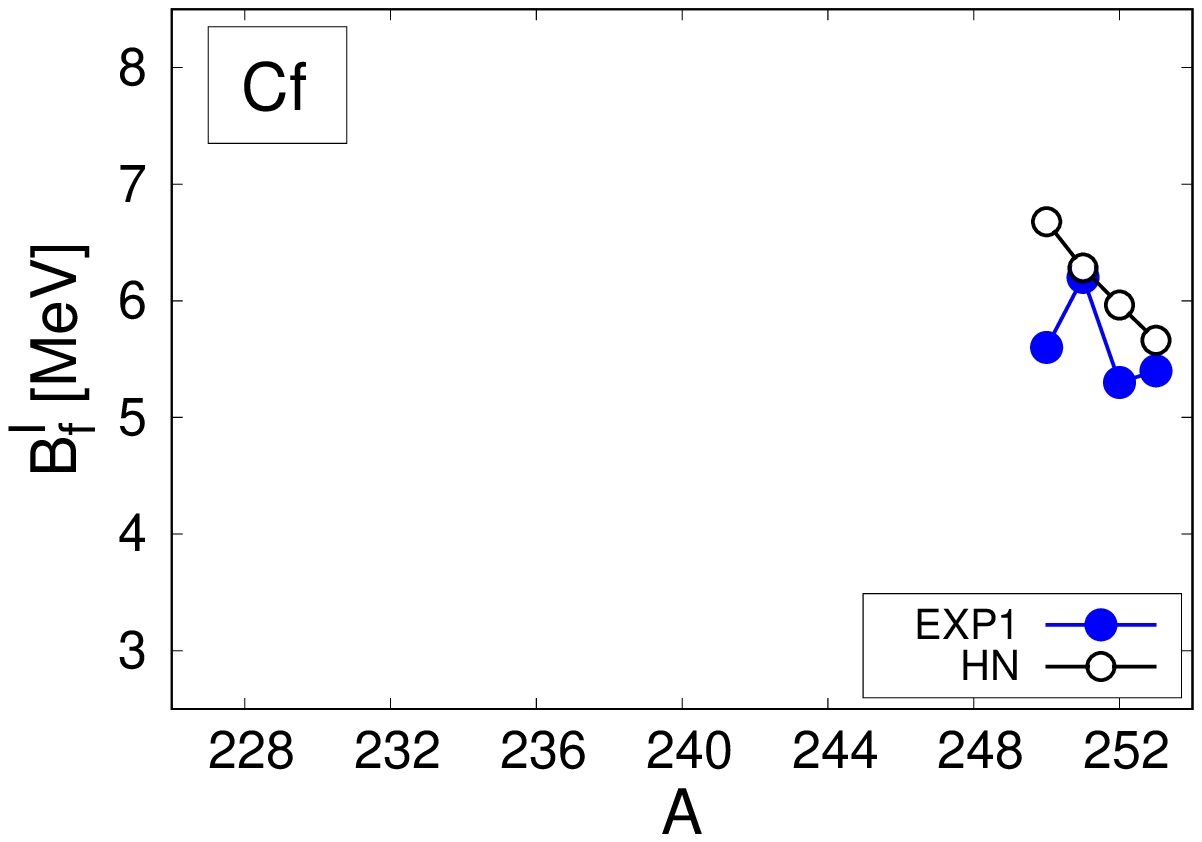}
\caption{
Calculated first fission-barrier heights HN (black circles) for
different isotopic chains compared with two sets of experimental data:
EXP1 \cite{Smirenkin1993} (red dots) and EXP2 \cite{Capote2009} (blue dots).}
\label{fig:BF1}
\end{figure*}

 The presently calculated (black circles) and experimental,
 EXP1 \cite{Smirenkin1993} and EXP2 \cite{Capote2009} (blue and red dots,
 respectively) first fission barrier heights $B_f^{I}$ are shown in
 Fig. \ref{fig:BF1}.
 Their numerical values are given in Table \ref{TABLE_TOT}, including results
 for \mbox{$^{226,227,228} {\rm Ac}$}. The latter nuclei will be discussed
 later (Sec. \ref{subsection:Thanomaly}) in more detail.

 The calculated barriers $B_f^{I}$ in Th nuclei are clearly too low compared
 to the experimental estimates. The difference is especially large in lighter
 isotopes. This discrepancy occurred in many other theoretical studies, eg.
\cite{Mamdouh1998,Samyn2005,Dobrowolski2007,Moller2009,Abusara2010},
 and will be discussed separately in Sec. \ref{subsection:Thanomaly}.
 Better agreement between calculated barriers and data occurs for protactinium
 and uranium isotopes for which our results lie quite close to,
 and sometimes between two sets of experimental points.
 In neptunium and plutonium nuclei our barriers become
 systematically higher than the empirical ones and they stay so in heavier
 actinides. The largest model vs experimental deviation can be observed in
 odd-odd americium isotopes, with discrepancies up to 1.6 MeV.
 With $Z$, the discrepancy between our results and data decreases in Cm,
 rises in Bk and becomes smaller again in Cf.

  Statistical parameters describing the deviation of calculated values of
 $B_f^{I}$ from the experimental estimates can be found in Table
 \ref{COMPARISONBF1}.
 Due to the lack of the empirical data for Ac isotopes, the comparison
 concerns nuclei from Th to Cf.
 In summary, the average discrepancy and the root mean square deviation do not
 exceed 1 MeV for both available sets of data.
 The inclusion of odd nuclei into consideration, without any tuning of
 parameters, worsens agreement with data compared the case of only even-even
 nuclei.

\begin{table}[h!]
\caption{Statistical parameters of the comparison of our
first fission barrier heights $B_f^{(I)th}$ with experimental estimates taken from \cite{Smirenkin1993,Capote2009}.
The average discrepancy $\bar{\Delta}$, and the rms deviation
$\delta_{\rm rms}$ are in MeV, where $N$ is the number
of considered nuclei.} \label{COMPARISONBF1}

\begin{tabular}{|c|c|c|}

\hline
\multicolumn{3}{ |c| }{comparison for $Z=90 \div 98$} \\
\hline
& $B_f^{(I)th}$ vs. EXP1 \cite{Smirenkin1993} & $B_f^{(I)th}$ vs. EXP2 \cite{Capote2009}  \\
$N$                   &    71       &    45        \\
$\bar{\Delta}$        &     0.80    &     0.73     \\
$\delta_{\rm rms}$    &     0.94    &     0.85     \\
\hline
\end{tabular}
\end{table}

Another observation concerns the odd-even staggering in barriers
which is definitely too strong compared to the data.
This effect was signalled in Sec. \ref{subsection:Groundstatemasses} and
 related to a too large decrease in the pairing gap due to blocking.
Sill, to better understand a source of this effect, in
 Sec. \ref{subsection:pairing} the role of the pairing interaction will be
additionally tested in selected cases.

\begin{table*}[t]
\scriptsize
\caption{Ground state masses: calculated $M_{gs}^{th}$ and measured $M_{gs}^{exp}$ \cite{Audi2003}.
Calculated first $B_f^{(I)th}$ and second $B_f^{(II)th}$ fission barrier heights compared with two sets of empirical compilations: EXP1 \cite{Smirenkin1993} and EXP2 \cite{Capote2009}, Excitation energy of the SD minimum - $E^{*th}_{II}$ relative to the ground state, experimental values of $E^{*exp}_{II}$ are taken from \cite{Singh1996}}
 \label{TABLE_TOT}
\centering
\begin{tabular}{|ccc|cc|ccc|cc|ccc|}
\hline
\multicolumn{3}{|c|}  {Nucleus} &
\multicolumn{2}{c|}  {} &
\multicolumn{3}{c|} {} &
\multicolumn{2}{c|} {} &
\multicolumn{3}{c|}  {} \\
\hline
$Z$ & $N$ & $A$ & $M_{gs}^{th}$ & $M_{gs}^{exp}$ & $B_f^{(I)th}$ & $B_f^{(I)EXP1}$ & $B_f^{(I)EXP2}$ & $E^{*th}_{II}$ & $E^{*exp}_{II}$ & $B_f^{(II)th}$ & $B_f^{(II)EXP1}$ & $B_f^{(II)EXP2}$ \\
\hline
89 & 137 & 226 & 23.58 &  24.31 &       4.07 & -   &  -        &  3.05  &  -                   & 7.16 & -   & 7.8  \\
89 & 138 & 227 & 25.48 &  25.85 &       3.94 & -   &  -        &  2.78  &  -                   & 6.96 & -   & 7.4  \\
89 & 139 & 228 & 28.25 &  28.90 &       4.38 & -   &  -        &  3.01  &  -                   & 6.80 & -   & 7.1  \\
\hline
90 & 137 & 227 & 25.15 &  25.81 &       3.74 & 5.9 &  -        &  2.87  &  -                   & 6.30 & 6.6 &  -   \\
90 & 138 & 228 & 26.62 &  26.77 &       3.57 & 6.2 &  -        &  2.48  &  -                   & 6.14 & 6.5 &  -   \\
90 & 139 & 229 & 29.06 &  29.59 &       4.17 & 5.9 &  -        &  2.90  &  -                   & 6.13 & 6.3 &  -   \\
90 & 140 & 230 & 30.71 &  30.86 &       3.98 & 6.1 & 6.1       &  2.62  &  -                   & 6.17 & 6.1 & 6.8  \\
90 & 141 & 231 & 33.33 &  33.82 &       4.78 & 6.0 & 6.0       &  2.35  &  -                   & 6.34 & 6.1 & 6.7  \\
90 & 142 & 232 & 35.31 &  35.45 &       4.55 & 5.8 & 5.8       &  2.11  &  -                   & 6.33 & 6.2 & 6.7  \\
90 & 143 & 233 & 38.36 &  38.73 &       5.21 & 6.1 & 5.1       &  1.49  &  $1.85 (\pm 0.25)$   & 6.35 & 6.3 & 6.65 \\
90 & 144 & 234 & 40.47 &  40.61 &       5.03 & 6.1 &  -        &  1.62  &  -                   & 6.33 & 6.3 &  -   \\
\hline
91 & 139 & 230 & 31.01 &  32.17 &       5.10 & 5.4 & 5.6       &  3.91  &  -                   & 6.81 & 5.4 & 5.8  \\
91 & 140 & 231 & 32.60 &  33.43 &       4.98 & 5.7 & 5.5       &  3.66  &  -                   & 6.91 & 5.7 & 5.5  \\
91 & 141 & 232 & 34.85 &  35.95 &       5.72 & 6.0 & 5.0       &  3.44  &  -                   & 7.05 & 6.1 & 6.4  \\
91 & 142 & 233 & 36.80 &  37.49 &       5.54 & 6.0 & 5.7       &  3.13  &  -                   & 6.95 & 6.0 & 5.8  \\
91 & 143 & 234 & 39.55 &  40.34 &       6.23 &  -  & 6.3       &  2.45  &  -                   & 6.87 & -   & 6.15 \\
\hline
92 & 139 & 231 & 33.07 &  33.81 &       4.64 & 5.2 & 4.4       &  3.41  &  -                   & 5.84 & 5.2 & 5.5  \\
92 & 140 & 232 & 34.30 &  34.61 &       4.52 & 5.4 & 4.9       &  3.10  &  -                   & 5.95 & 5.3 & 5.4  \\
92 & 141 & 233 & 36.40 &  36.92 &       5.29 & 5.7 & 4.35      &  2.86  &  -                   & 6.23 & 5.7 & 5.55 \\
92 & 142 & 234 & 37.98 &  38.15 &       5.12 & 5.9 & 4.8       &  2.57  &  -                   & 6.16 & 5.7 & 5.5  \\
92 & 143 & 235 & 40.57 &  40.92 &       5.86 & 6.0 & 5.25      &  1.94  &  -                   & 6.14 & 5.8 & 6.0  \\
92 & 144 & 236 & 42.28 &  42.45 &       5.69 & 5.6 & 5.0       &  2.05  &  $2.75 (\pm 0.01)$   & 6.13 & 5.6 & 5.67 \\
92 & 145 & 237 & 45.04 &  45.39 &       6.45 & 6.2 & 6.4       &  1.92  &  -                   & 6.49 & 5.9 & 6.15 \\
92 & 146 & 238 & 47.15 &  47.31 &       6.06 & 6.0 & 6.3       &  1.94  &  2.56                & 6.27 & 5.8 & 5.5  \\
92 & 147 & 239 & 50.23 &  50.57 &       6.70 & 6.3 & 6.45      &  2.02  &  -                   & 7.05 & 6.0 & 6.0  \\
92 & 148 & 240 & 52.66 &  52.72 &       6.13 & 6.1 &  -        &  2.04  &  -                   & 6.59 & 5.8 &  -   \\
\hline
93 & 140 & 233 & 37.53 &  37.95 &       5.14 & 5.0 &  -        &  3.46  &  -                   & 5.86 & 5.1 &  -   \\
93 & 141 & 234 & 39.25 &  39.96 &       6.10 & 5.5 &  -        &  3.31  &  -                   & 6.35 & 5.4 &  -   \\
93 & 142 & 235 & 40.71 &  41.04 &       5.89 & 5.5 &  -        &  3.06  &  -                   & 6.24 & 5.5 &  -   \\
93 & 143 & 236 & 42.89 &  43.38 &       6.79 & 5.8 & 5.9       &  2.58  &  -                   & 6.40 & 5.6 & 5.4  \\
93 & 144 & 237 & 44.48 &  44.87 &       6.54 & 5.7 & 6.0       &  2.69  &  $2.80 (\pm 0.40)$   & 6.44 & 5.5 & 5.4  \\
93 & 145 & 238 & 46.70 &  47.46 &       7.41 & 6.0 & 6.5       &  2.67  &  -                   & 6.98 & 5.9 & 5.75 \\
93 & 146 & 239 & 48.87 &  49.31 &       6.98 & 5.8 &  -        &  2.56  &  -                   & 6.60 & 5.4 &  -   \\
\hline
94 & 141 & 235 & 41.78 &  42.18 &       5.64 & 5.7 &  -        &  2.64  &  $3.00 (\pm 0.20)$   & 5.37 & 5.1 &  -   \\
94 & 142 & 236 & 42.88 &  42.90 &       5.49 & 5.7 &  -        &  2.42  & $\sim 3.00$          & 5.32 & 4.5 &  -   \\
94 & 143 & 237 & 44.95 &  45.09 &       6.26 & 5.6 & 5.10      &  1.92  & $ 2.60 (\pm 0.20)$   & 5.48 & 5.4 & 5.15 \\
94 & 144 & 238 & 46.16 &  46.16 &       6.24 & 5.9 & 5.6       &  2.04  & $\sim 2.40$          & 5.55 & 5.2 & 5.1  \\
94 & 145 & 239 & 48.31 &  48.59 &       7.08 & 6.2 & 6.2       &  2.02  & $ 3.10 (\pm 0.20)$   & 6.01 & 5.5 & 5.7  \\
94 & 146 & 240 & 50.06 &  50.13 &       6.61 & 5.8 & 6.05      &  1.94  & $\sim 2.80$          & 5.71 & 5.3 & 5.15 \\
94 & 147 & 241 & 52.66 &  52.96 &       7.08 & 6.2 & 6.15      &  1.94  & $\sim 2.20$          & 6.53 & 5.6 & 5.50 \\
94 & 148 & 242 & 54.65 &  54.72 &       6.60 & 5.7 & 5.85      &  1.97  & $\sim 2.20$          & 6.09 & 5.3 & 5.05 \\
94 & 149 & 243 & 57.66 &  57.76 &       6.70 & 5.9 & 6.05      &  2.17  & $ 1.70 (\pm 0.30)$   & 6.80 & 5.5 & 5.45 \\
94 & 150 & 244 & 59.80 &  59.81 &       6.37 & 5.5 & 5.7       &  2.14  &  -                   & 6.35 & 5.2 & 4.85 \\
94 & 151 & 245 & 62.88 &  63.11 &       6.58 & 5.5 & 5.85      &  2.81  & $ 2.00 (\pm 0.40)$   & 7.13 & 5.4 & 5.25 \\
94 & 152 & 246 & 65.43 &  65.40 &       6.02 & 5.4 &  -        &  2.44  &  -                   & 6.50 & 5.3 &  -   \\
\hline
95 & 144 & 239 & 49.42 &  49.39 &       6.94 & 6.3 & 6.00      &  2.19  & $ 2.50 (\pm 0.20)$   & 5.44 & 4.9 & 5.40 \\
95 & 145 & 240 & 51.17 &  51.51 &       7.72 & 6.4 & 6.10      &  2.19  & $ 3.00 (\pm 0.20)$   & 6.00 & 5.2 & 6.00 \\
95 & 146 & 241 & 52.81 &  52.94 &       7.46 & 6.2 & 6.00      &  2.10  & $\sim 2.20$          & 5.63 & 5.1 & 5.35 \\
95 & 147 & 242 & 55.02 &  55.47 &       7.82 & 6.4 & 6.32      &  2.02  & $ 2.20 (\pm 0.08)$   & 6.57 & 5.4 & 5.78 \\
95 & 148 & 243 & 56.94 &  57.18 &       7.31 & 6.1 & 6.40      &  2.07  & $ 2.30 (\pm 0.20)$   & 6.09 & 5.4 & 5.05 \\
95 & 149 & 244 & 59.59 &  59.88 &       7.44 & 6.2 & 6.25      &  2.41  & $ 2.80 (\pm 0.40)$   & 6.68 & 5.4 & 5.9  \\
95 & 150 & 245 & 61.66 &  61.90 &       6.93 & 6.1 &  -        &  2.23  & $ 2.40 (\pm 0.40)$   & 6.23 & 5.2 &  -   \\
95 & 151 & 246 & 64.42 &  64.99 &       7.02 & 5.8 &  -        &  2.86  & $\sim 2.00$          & 6.98 & 5.0 &  -   \\
95 & 152 & 247 & 66.92 & (67.15)&       6.56 & 5.7 &  -        &  2.43  &  -                   & 6.26 & 4.8 &  -   \\
\hline
96 & 145 & 241 & 53.65 &  53.70 &       7.33 & 6.4 & 7.15      &  1.65  & $\sim 2.30$          & 5.14 & 4.3 & 5.5  \\
96 & 146 & 242 & 54.88 &  54.81 &       6.96 & 6.0 & 6.65      &  1.64  & $ 1.90 (\pm 0.20)$   & 4.85 & 4.0 & 5.0  \\
96 & 147 & 243 & 56.99 &  57.18 &       7.34 & 6.5 & 6.33      &  1.57  & $ 1.90 (\pm 0.30)$   & 5.76 & 4.6 & 5.4  \\
96 & 148 & 244 & 58.51 &  58.45 &       6.91 & 6.1 & 6.18      &  1.66  & $\sim 2.20$          & 5.36 & 4.3 & 5.10 \\
96 & 149 & 245 & 61.01 &  61.00 &       7.10 & 6.3 & 6.35      &  1.97  & $ 2.10 (\pm 0.30)$   & 6.04 & 4.9 & 5.45 \\
96 & 150 & 246 & 62.72 &  62.62 &       6.68 & 6.0 & 6.0       &  1.89  &  -                   & 5.63 & 4.7 & 4.80 \\
96 & 151 & 247 & 65.29 &  65.53 &       6.98 & 6.1 & 6.12      &  2.60  &  -                   & 6.53 & 4.9 & 5.10 \\
96 & 152 & 248 & 67.44 &  67.39 &       6.38 & 5.9 & 5.8       &  2.24  &  -                   & 5.89 & 5.0 & 4.80 \\
96 & 153 & 249 & 70.94 &  70.75 &       6.02 & 5.7 & 5.63      &  2.20  &  -                   & 5.83 & 4.7 & 4.95 \\
96 & 154 & 250 & 73.04 &  72.99 &       5.72 & 5.4 &  -        &  2.12  &  -                   & 5.52 & 4.4 &  -   \\
\hline
97 & 147 & 244 & 60.36 &  60.72 &       7.68 & 6.6 &  -        &  1.26  &  -                   & 5.42 & 4.2 &  -   \\
97 & 148 & 245 & 61.79 &  61.82 &       7.19 & 6.4 &  -        &  1.37  & $\sim 1.56$          & 5.07 & 4.2 &  -   \\
97 & 149 & 246 & 63.88 &  63.97 &       7.40 & 6.5 &  -        &  1.85  &  -                   & 5.78 & 4.7 &  -   \\
97 & 150 & 247 & 65.51 &  65.49 &       7.02 & 6.5 &  -        &  1.66  &  -                   & 5.38 & 4.6 &  -   \\
97 & 151 & 248 & 67.66 & (68.08)&       7.49 & 6.3 &  -        &  2.31  &  -                   & 6.22 & 4.8 &  -   \\
97 & 152 & 249 & 69.77 &  69.85 &       6.77 & 6.1 &  -        &  1.98  &  -                   & 5.53 & 4.5 &  -   \\
97 & 153 & 250 & 72.92 &  72.95 &       6.35 & 6.1 &  -        &  1.69  &  -                   & 5.04 & 4.1 &  -   \\
\hline
98 & 152 & 250 & 71.28 &  71.17 &       6.67 & 5.6 &  -        &  1.83  &  -                   & 5.14 & 3.8 &  -   \\
98 & 153 & 251 & 74.35 &  74.13 &       6.25 & 6.2 &  -        &  1.63  &  -                   & 4.58 & 3.9 &  -   \\
98 & 154 & 252 & 76.08 &  76.03 &       5.97 & 5.3 &  -        &  1.58  &  -                   & 4.21 & 3.5 &  -   \\
98 & 155 & 253 & 79.41 &  79.30 &       5.61 & 5.4 &  -        &  1.06  &  -                   & 3.59 & 3.5 &  -   \\
\hline
\end{tabular}
\end{table*}

\subsection{Second fission barrier heights}

\begin{figure*}[h]
\includegraphics[scale=0.4]{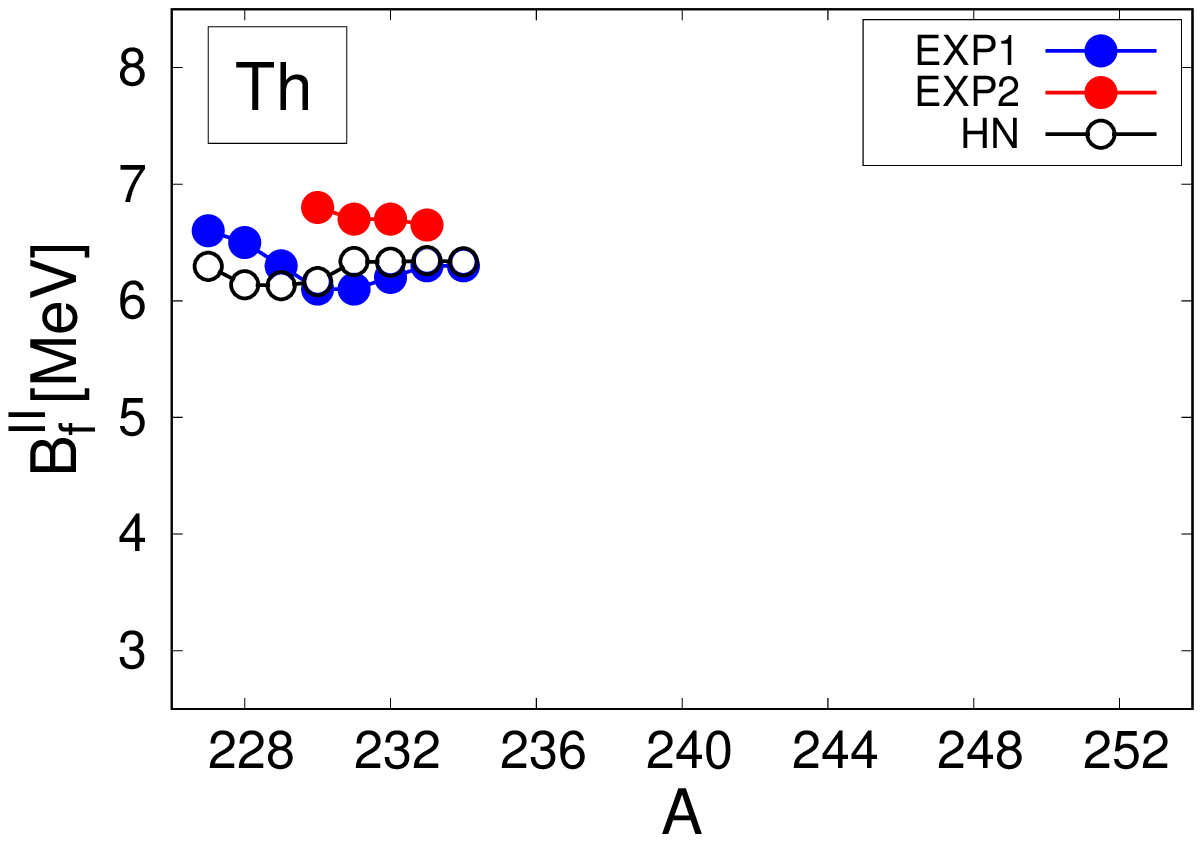}
\includegraphics[scale=0.4]{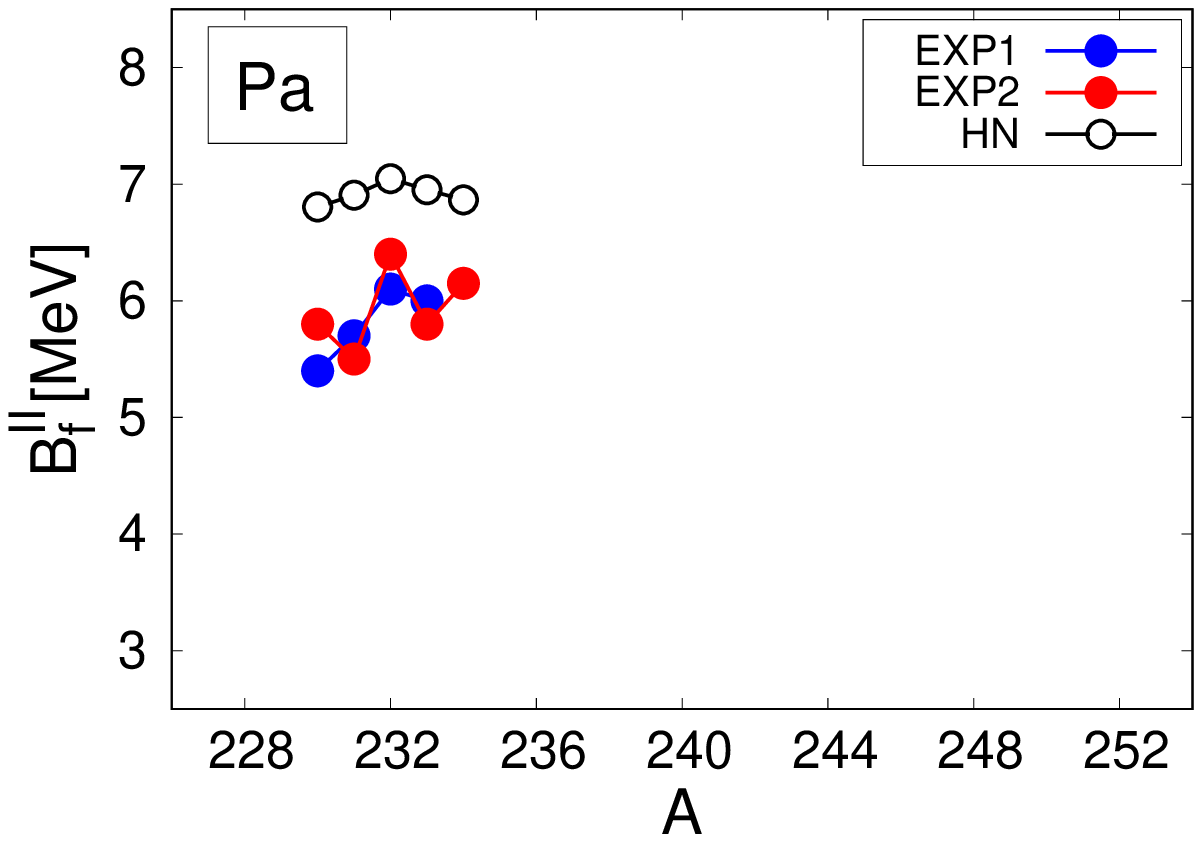}
\includegraphics[scale=0.4]{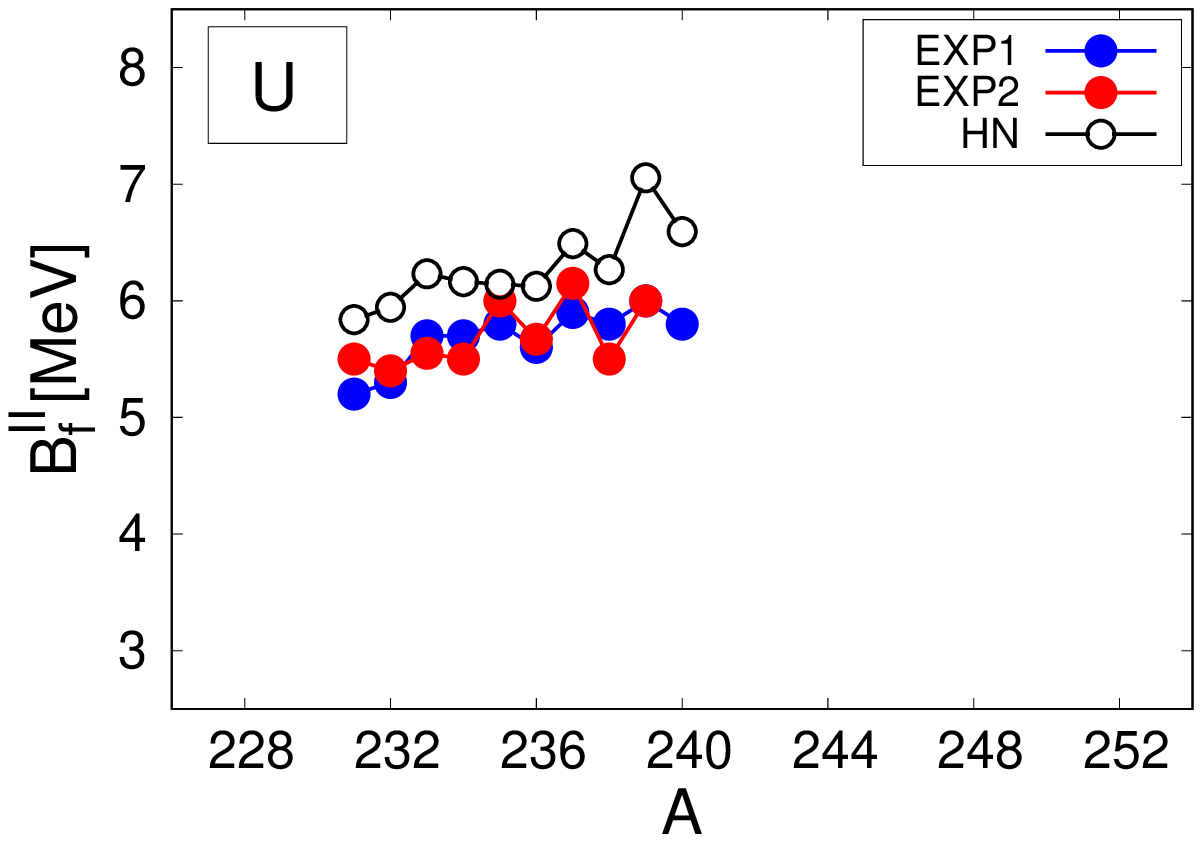}
\includegraphics[scale=0.4]{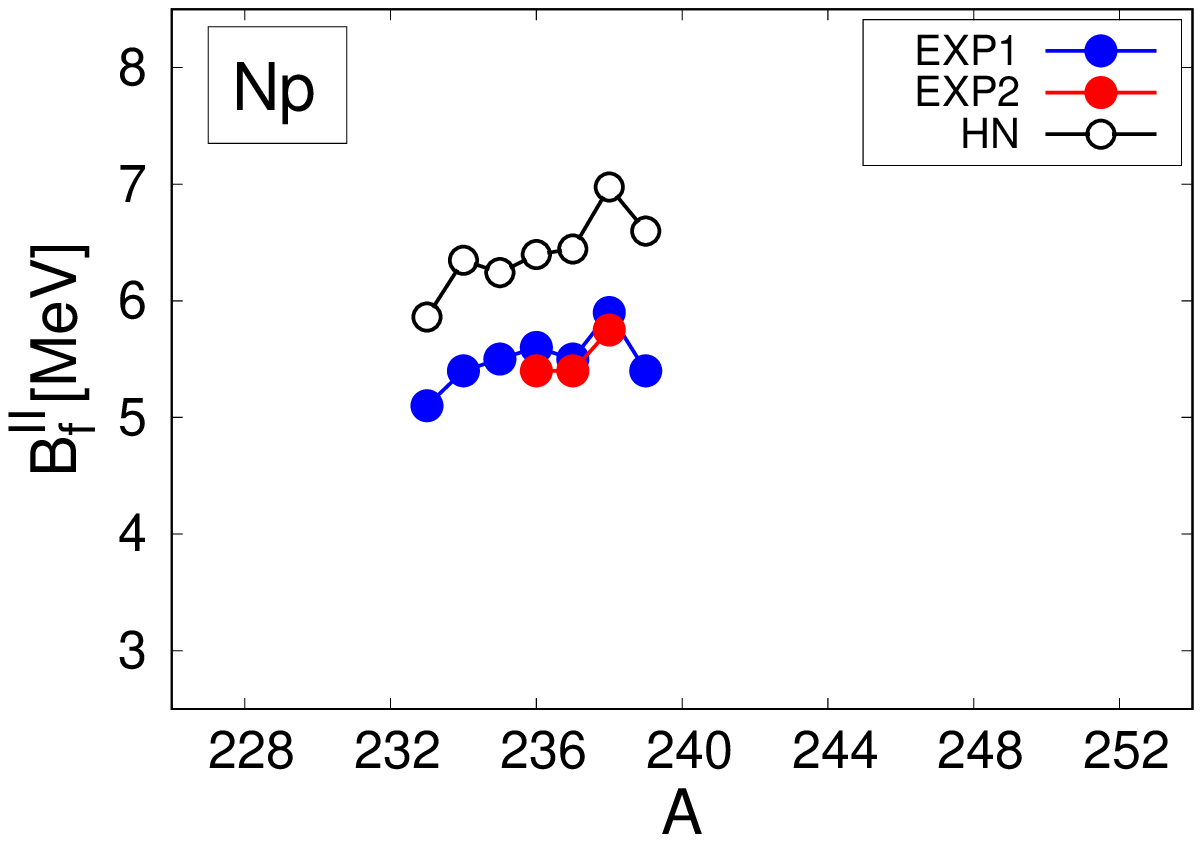}
\includegraphics[scale=0.4]{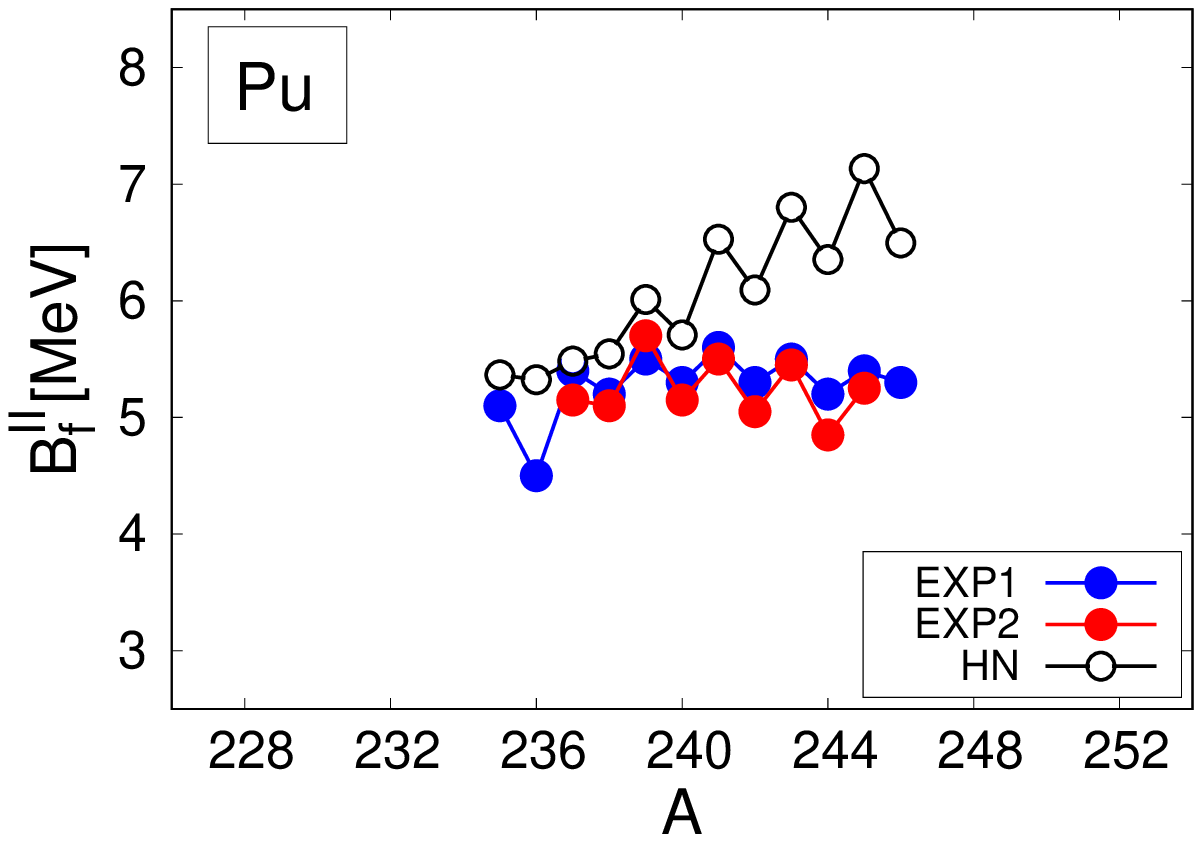}
\includegraphics[scale=0.4]{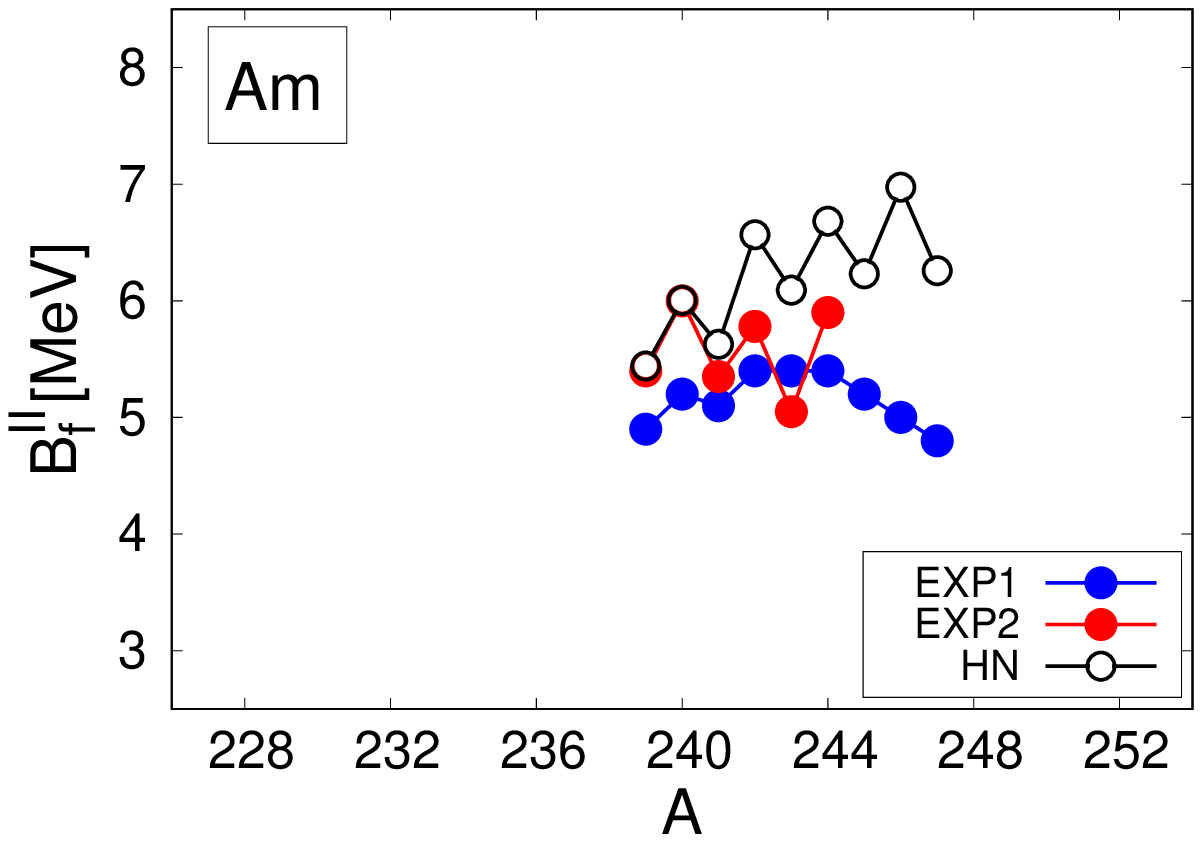}
\includegraphics[scale=0.4]{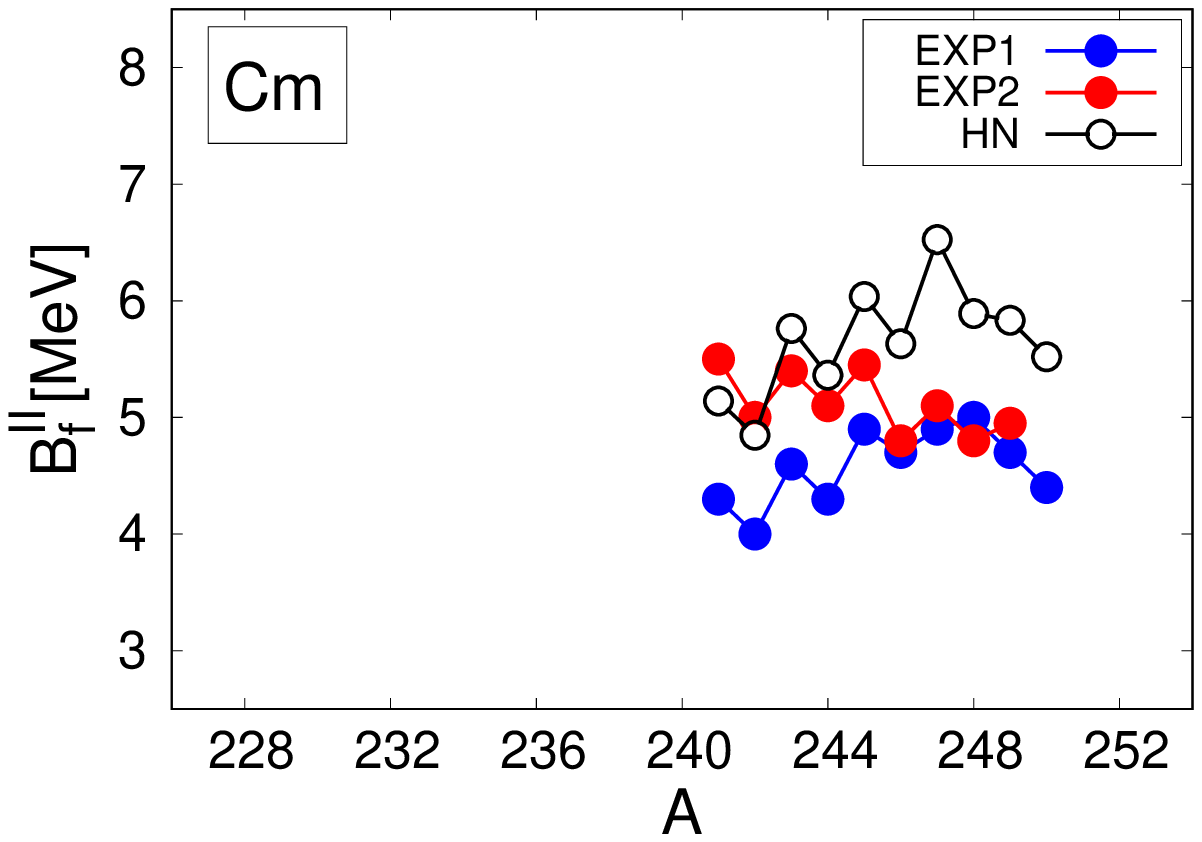}
\includegraphics[scale=0.4]{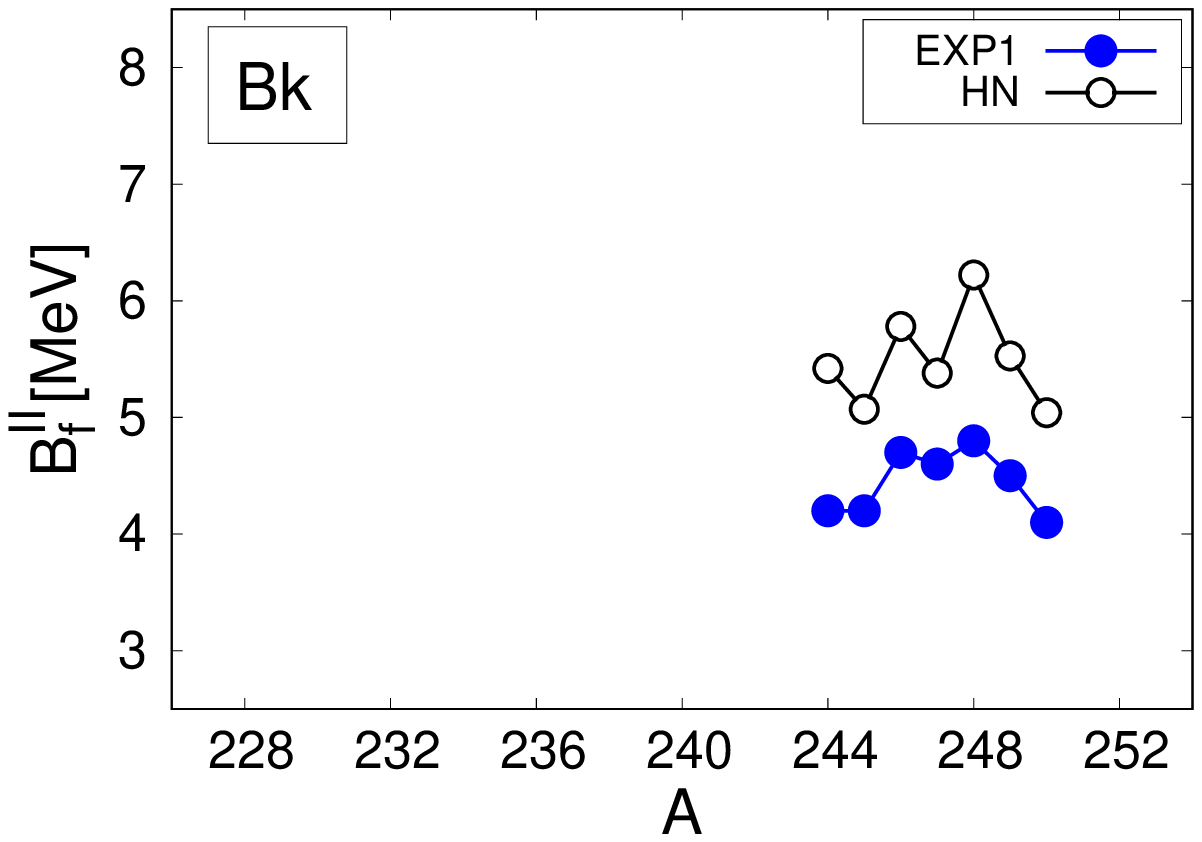}
\includegraphics[scale=0.4]{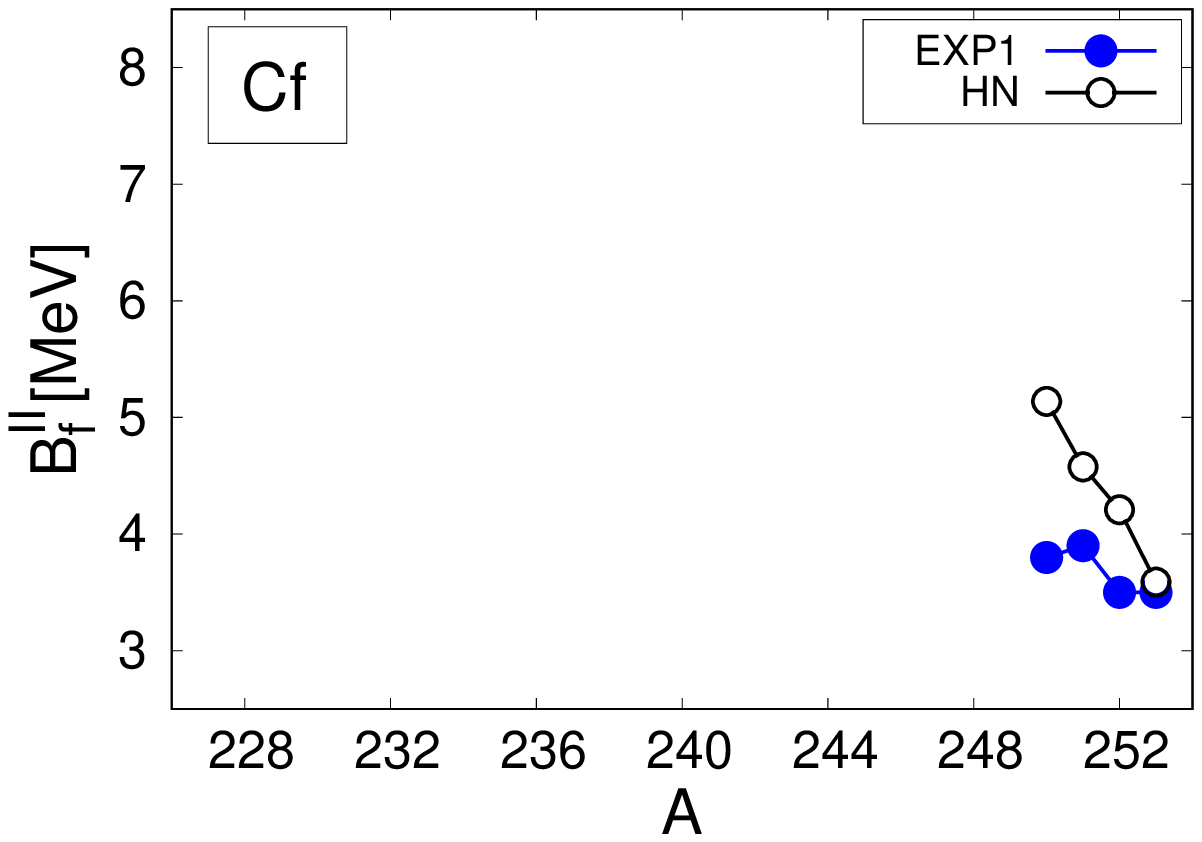}
\caption{
Calculated second fission-barrier heights HN (black circles)
for different isotopic chains compared
with two sets of experimental data: EXP1 \cite{Smirenkin1993} (red dots)
and EXP2 \cite{Capote2009} (blue dots).}
\label{fig:BF2}
\end{figure*}

 A comparison between experimental and calculated
 second barrier heights $B_f^{II}$ is presented in Fig. \ref{fig:BF2} as well
 as in the last columns of Table \ref{TABLE_TOT}.
 It should be emphasized that the two sets of experimental data for second
 fission barriers differ more than 0.5 MeV in Th and Cm;
 for example, in $^{242}$Cm this difference amounts to ~ 1 MeV.
  They also differ in a subtle way:
  the Am data taken from \cite{Capote2009} indicate a quite strong
 odd - even staggering while those from \cite{Smirenkin1993} do not.
 In Cm nuclei the odd-even staggering for both experimental data sets is
 already similar.

 As one can see, our $B_f^{II}$ are almost always higher than the experimental
 ones. In uranium and neptunium isotopes the general trend of
the experimental data seems to be reproduced quite well.
 The largest discrepancy of 1.5 - 2 MeV between calculated and experimental
 barriers occurs for odd-odd americium isotopes (as for the first barriers)
 and for odd-neutron Pu and Cm chains.

There are also discrepancies suggesting more involved reasons. In Pu and
Am isotopes the barriers increase with $N$ while no such effect is
 observed in the data. A similar increase was also produced in other
 micro-macro \cite{Moller2009, Dobrowolski2007} and non-relativistic self-consistent calculations - see \cite{Delaroche2006,Robledo2014}
 and \cite{BENDER2004} (in Fig. 3, for Sly6 interaction). This problem seems to be absent
 in the RMF approach, see \cite{Zhou2014} and \cite{BENDER2004} for NL-Z2 and NL3 models.

It is worth noting that for light actinides the odd - even staggering in
 second barriers is practically absent.
 It becomes more pronounced for the mass numbers greater than 238, and is
 clearly visible in plutonium isotopes, as well as in the heavier isotopic
 chains.
 Then, it disappears for the neutron numbers greater than 152.

 The mean discrepancy and rms deviation of the second barriers $B_f^{II}$ can
 be found in Table. \ref{COMPARISONBF2}.
 Comparison was done starting from Ac and ending at Cf nuclei.
 As for the first barriers before, the statistical deviations between our
 second fission barriers and data are less than 1 MeV.
 For the even-even systems, the present barriers can be compared with our
 previous results \cite{Jachimowicz2012}.
 Despite the fact that the currently used method is slightly different in
 including the dipole deformation $\beta_{10}$,
  the second barriers stay as they were.

\begin{table}[h!]
\caption{The same as in Tab. \ref{COMPARISONBF1}
but for our second fission barrier heights $B_f^{(II)th}$.} \label{COMPARISONBF2}
\begin{tabular}{|c|c|c|}
\hline
\multicolumn{3}{ |c| }{comparison for $Z=89 \div 98$} \\
\hline
& $B_f^{(II)th}$ vs. EXP1 \cite{Smirenkin1993} & $B_f^{(II)th}$ vs. EXP2 \cite{Capote2009}  \\
$N$                   &    71       &    48        \\
$\bar{\Delta}$        &     0.82    &     0.70     \\
$\delta_{\rm rms}$    &     0.92    &     0.82     \\
\hline
\end{tabular}
\end{table}

\subsection{Fission barriers of Ac and Th isotopes} \label{subsection:Thanomaly}

 As mentioned before, the calculated first fission barriers ($B_f^{I}$) in light
 Th nuclei are significantly lower than the second ones
 ($B_f^{II}$) and, at the same time, much smaller than the
 experimental first barriers - see \mbox{Table \ref{TABLE_TOT}}.
 For example, in $^{228}$Th the latter difference is greater than 2 MeV.
 Curiously, the three experimental inner barriers in $^{227 - 229}$Th show a
 reversed odd-even staggering, with the highest barrier in the even - even
 isotope.

To study the intriguing puzzle of too-low first fission barriers
in light thorium nuclei, the so called "{\it thorium anomaly}",
 we turned to comparisons of PES's obtained for Th with those for slightly
 lighter Ac isotopes.
 According to the empirical data, the outer fission barrier in (pre-)actinides
 is much higher than the first (inner) one \cite{Nemilov1983} and this is
 probably why in many experimental studies the single-humped barrier is
  considered as the correct one, cf \mbox{Table \ref{TABLE_TOT}}
 for $^{226-228}\rm {Ac}$, where the single experimental barriers are
 close to  our calculated second barriers rather than the first.
 In Fig. \ref{fig:Ac_Th}, we show energy surfaces for $^{227}$Ac and
 $^{228}$Th obtained, as other PES's presented below, from the
  7-dimensional grid (\ref{sp2grid}) by the energy minimization over 5 not
 displayed deformations (with dipole deformation $\beta_{10}$ fixed by the
 center of mass condition).
 The ground state in $^{227}$Ac is calculated at $\beta_{20} \approx 0.20$,
 the second minimum at $\beta_{20} \approx 0.50$, and a very shallow third
 minimum at $\beta_{20}\approx 1.00,\beta_{30}\approx 0.25$.
 As can be seen in the map, the second fission barrier is much higher and more
 elongated than the rather unprominent first one. It should be also kept
 in mind that the first barrier is still reduced by the triaxiality, not
 included in Fig. \ref{fig:Ac_Th}.
 The PES for $^{228} {\rm Th}$, also in \mbox{Fig. \ref{fig:Ac_Th}},
 is very similar to that of $^{227} {\rm Ac}$, and both second barriers
 are close to the experimental estimates. From the point of view of
 our results it would be natural to interpret barriers in
 both nuclei in the same manner. However, in the empirical interpretation
 there is no first barrier in $^{227}$Ac, while the one in $^{228}$Th
 is nearly as high (6.2 MeV) as the second one (6.5 MeV). Surely, it would be
 good to understand the reason of such an abrupt change.

 A sequence of four maps in Fig. \ref{fig:4Th} for odd-neutron
thorium isotopes illustrates the calculated evolution - i.e. heights and mutual
 positions - of the first and second fission saddles
  (the PES's for even-even thorium isotopes show very similar picture).
 As one can see, with increasing neutron number, the first fission barrier,
corresponding to $\beta_{20} \approx 0.30-0.40$, becomes more pronounced
 while initially shallow second minimum becomes more deep. For
 $^{231,233} {\rm Th}$, the second fission barrier splits into
 two peaks divided by a shallow third minimum about 0.5 MeV deep.
 For heavier actinides, the second barrier becomes much shorter and the first
 one becomes dominant.
 Two representative cases of such different energy landscapes are shown
 in \mbox{Fig. \ref{fig:Z_evolution}}: in $^{235}{\rm U}$, both calculated
 barrier heights $B_f^{I}$ and $B_f^{II}$ are similar, while in
 $^{251} \rm {Cf}$ the second peak nearly vanishes.
 This illustrates the evolution of both fission barriers in
  actinides with increasing number of protons.

 The calculated evolution of the barriers is not fully reflected
 in experimental evaluations. In particular, the curvature of the fission
 barrier (at the saddle point), on which the transmission coefficient depends
 exponentially, was assumed constant for groups of nuclei in
  \cite{Capote2009}, with the following values (in MeV):
\begin{center}
\begin{tabular}{cccc}
\hline
              & \multicolumn{3}{c}{nucleus:} \\
              & even-even & odd & odd-odd \\
 inner hump : & 0.9-1.0   & 0.8 & 0.6     \\
outer  hump : & 0.6       & 0.5 & 0.4     \\
\hline
\end{tabular}
\end{center}

 A conceptual difficulty in comparing calculated and experimental fission
  barriers is the multidimesionality, i.e. a multitude of deformation
 parameters involved in the fission process.
 While inherent in the PES approach, it is omitted in the  empirical estimates
  which are based on one-dimensional models.
  This can be clearly appreciated when viewing one of the maps, e.g.
   for $^{235}$U in \mbox{Fig. \ref{fig:Z_evolution}},
  where it may be seen that the curvature at the saddle will
 depend on the direction it is traversed - in this map it will be the
 choice of the direction in the quadrupole - octupole $(\beta_{20},\beta_{30})$
  plane, but in general it will involve all employed deformations.
 This relates to the nature of fission as a dynamic process, while
  the picture used here is static.

  Finally, the occurrence of the third minimum and the third barrier
  additionally complicates the description of fission.
 In these calculations, the third barriers in $^{227-229} {\rm Th}$ are
 smaller than 0.5 MeV, while for $^{231-233} {\rm Th}$ they are larger than
 0.5 MeV and visible in \mbox{Fig.\ref{fig:4Th}}. One should note, however,
 that the third barriers come out lower when one allows for an independent
 change in $\beta_{10}$ (i.e. when $\beta_{10}$ is not fixed by the center
 of mass condition as here), as in \cite{Kowal2012,Jachimowicz2013}.

\begin{figure*}[h]
\includegraphics[scale=0.46]{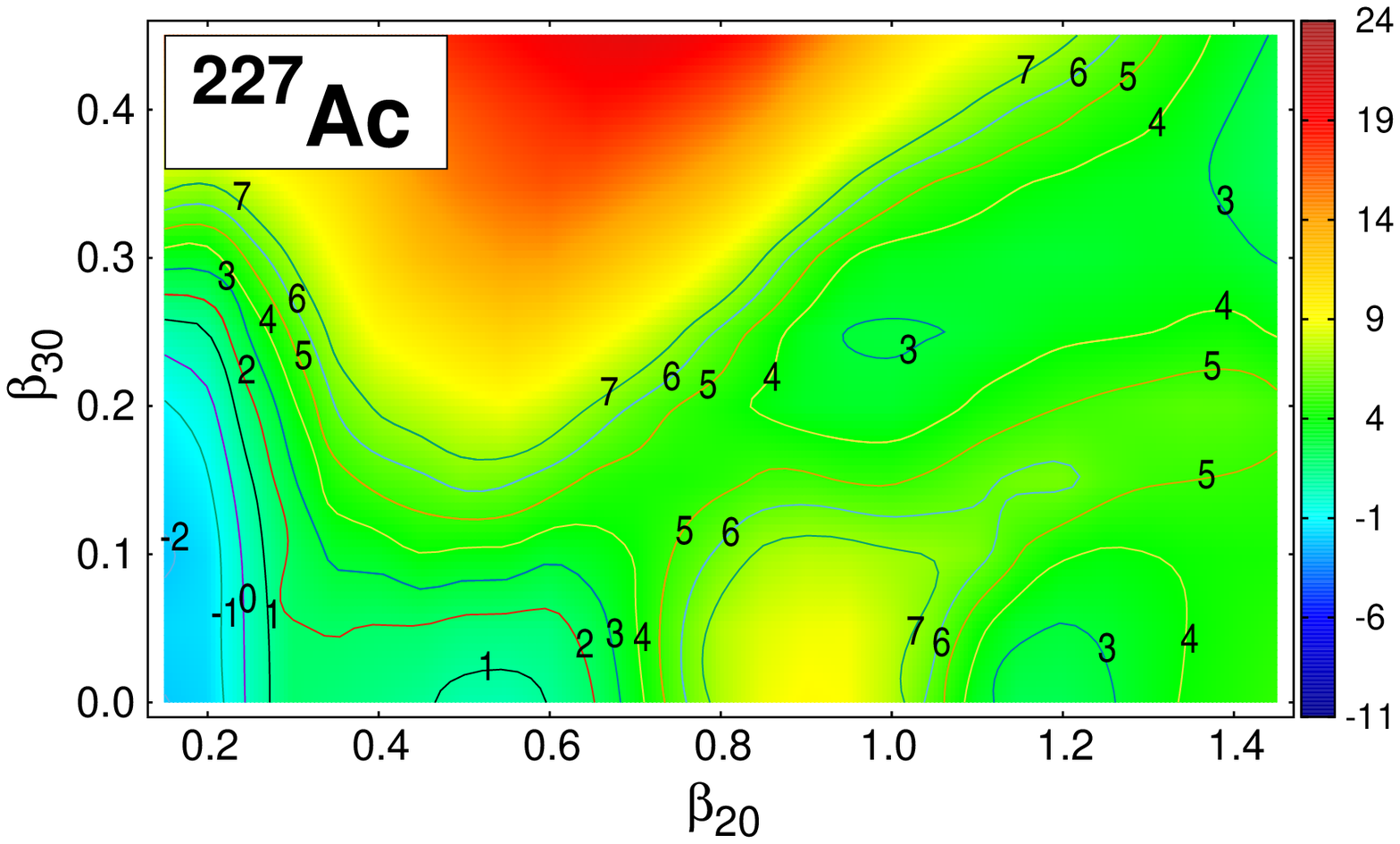}
\includegraphics[scale=0.46]{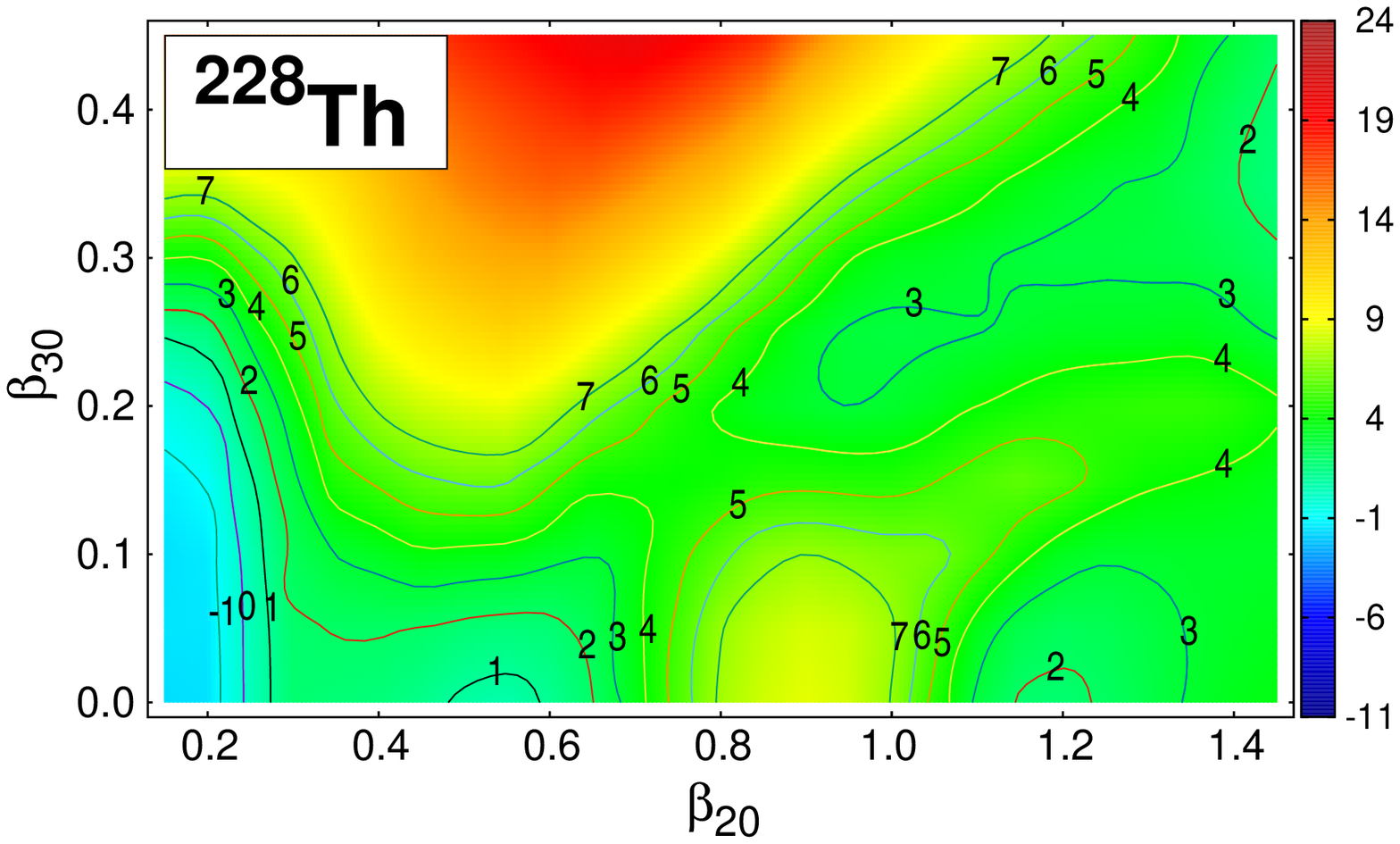}
\caption{Energy landscapes in $\{ \beta_{20}$, $\beta_{30} \}$ plane for
$^{227} {\rm Ac}$ and $^{228} {\rm Th}$,
calculated from the minimization over the remaining five
deformation parameters: $\{ \beta_{40}$, $\beta_{50}$,
$\beta_{60}$, $\beta_{70}$, $\beta_{80} \}$, where $\beta_{10}$ was fixed by the center of
mass condition. Energy (in MeV) calculated relative to the macroscopic energy at the spherical shape.}
\label{fig:Ac_Th}
\end{figure*}

\begin{figure*}[h]
\includegraphics[scale=0.46]{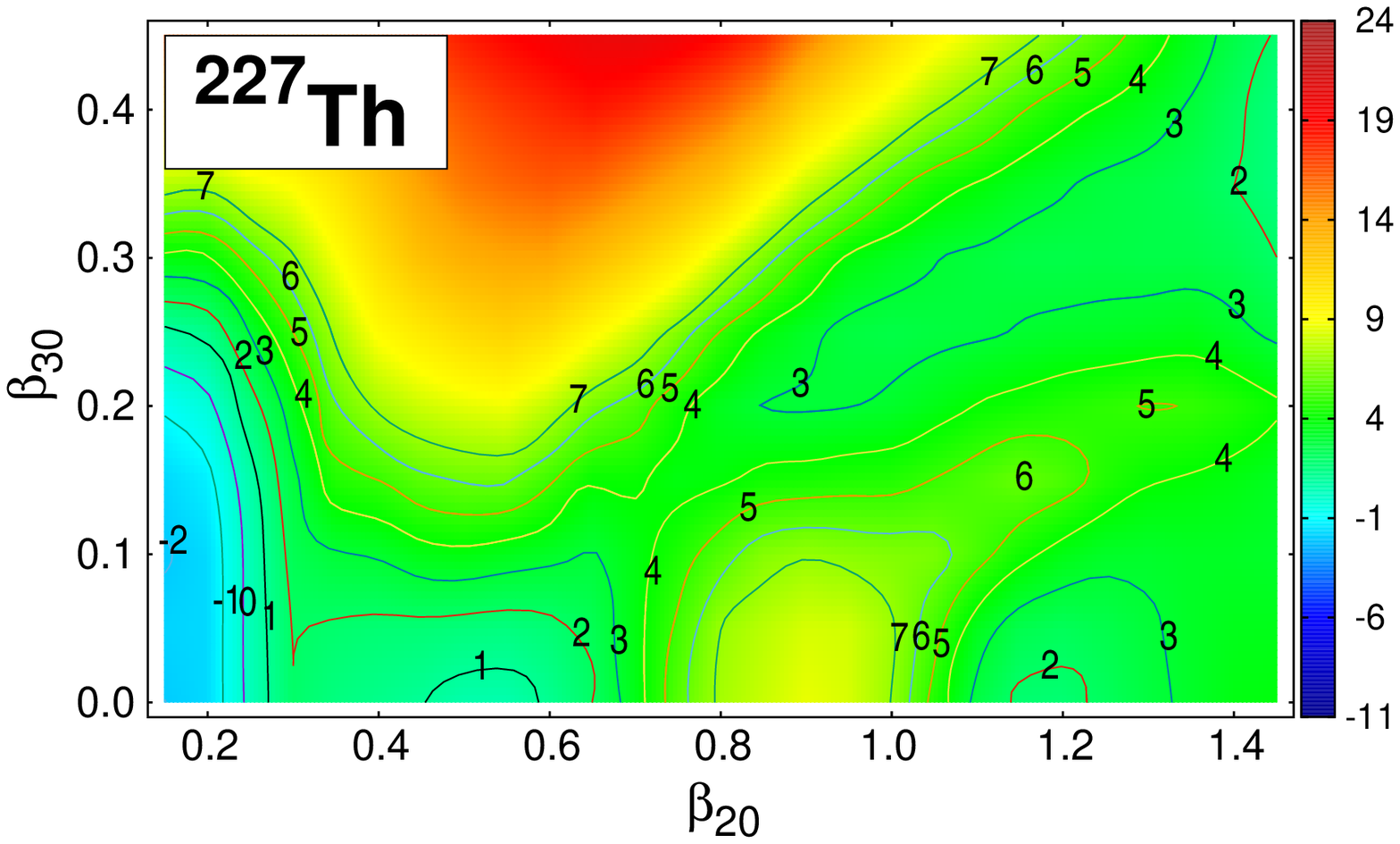}
\includegraphics[scale=0.46]{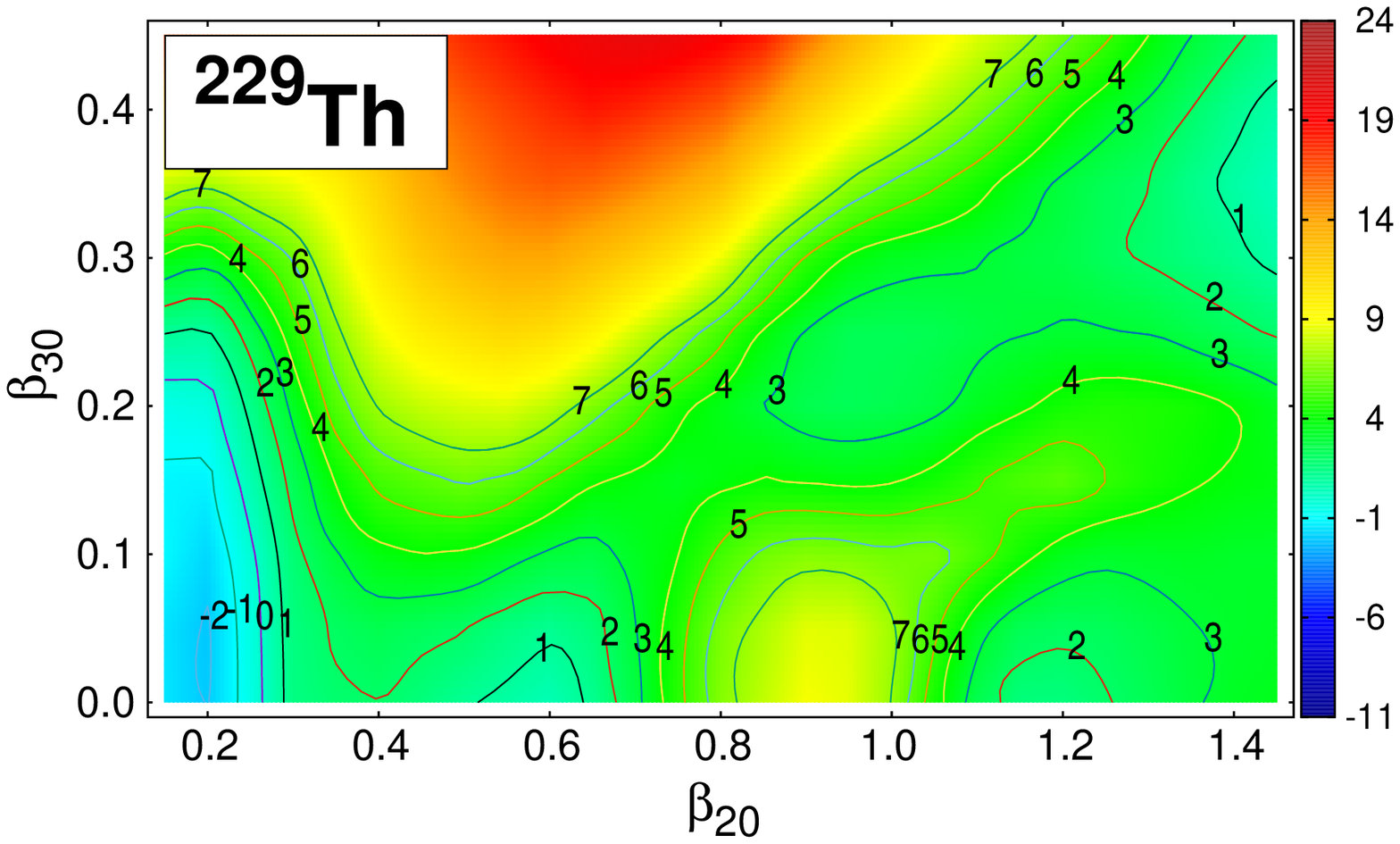}
\includegraphics[scale=0.46]{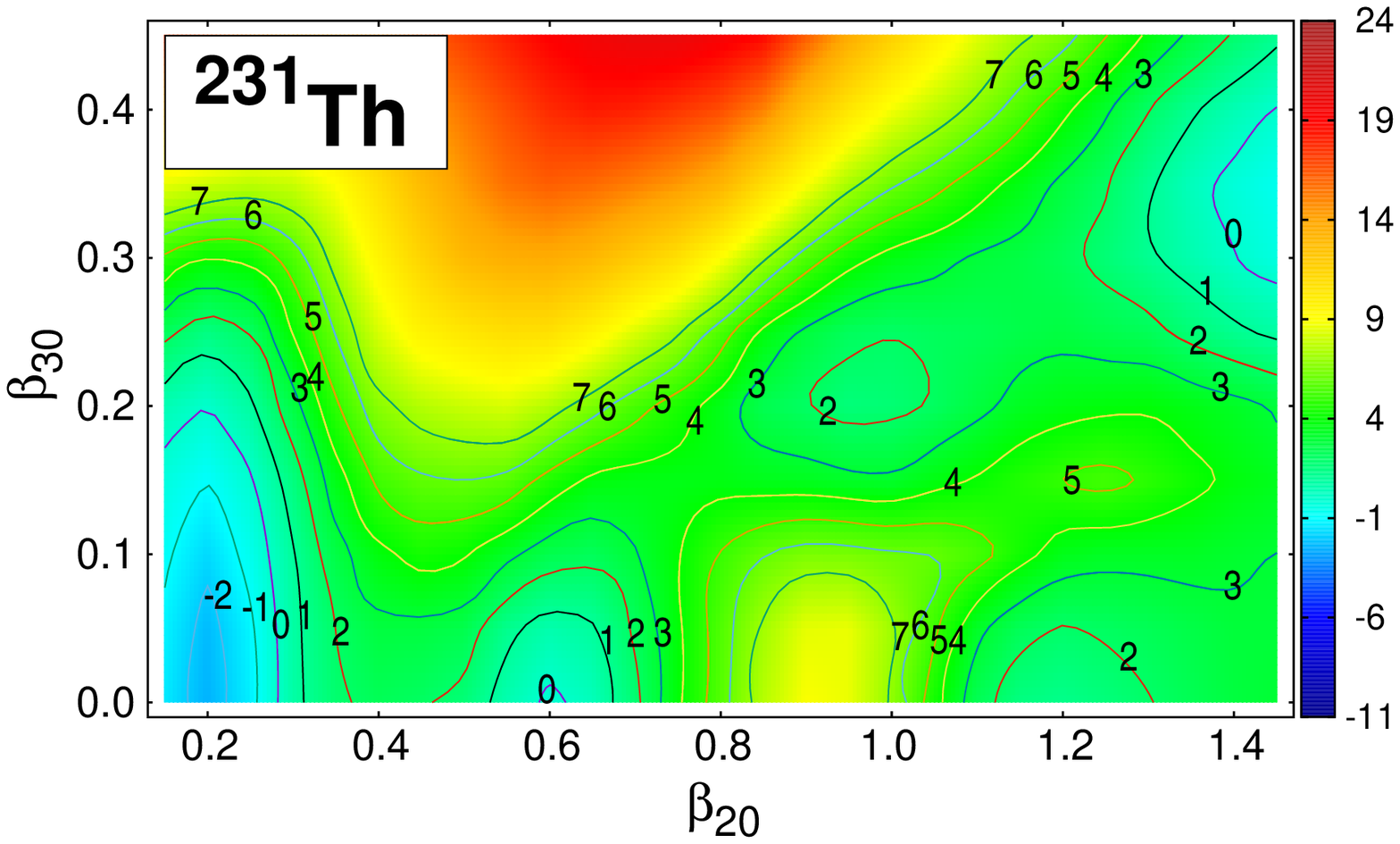}
\includegraphics[scale=0.46]{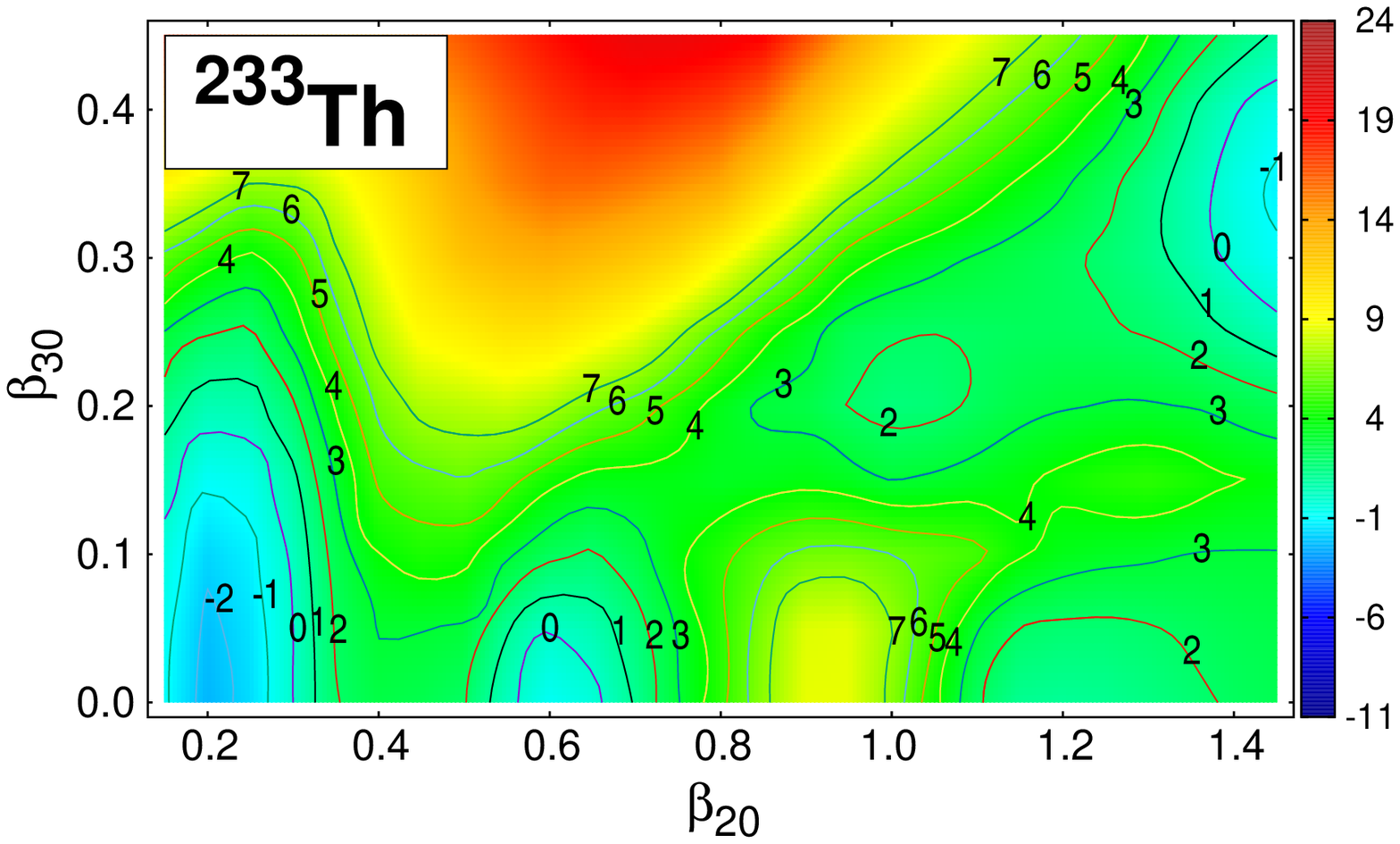}
\caption{The same as in Fig. \ref{fig:Ac_Th} but for $^{227,229,231,233} {\rm Th}$. }
\label{fig:4Th}
\end{figure*}

\begin{figure*}[h]
\includegraphics[scale=0.46]{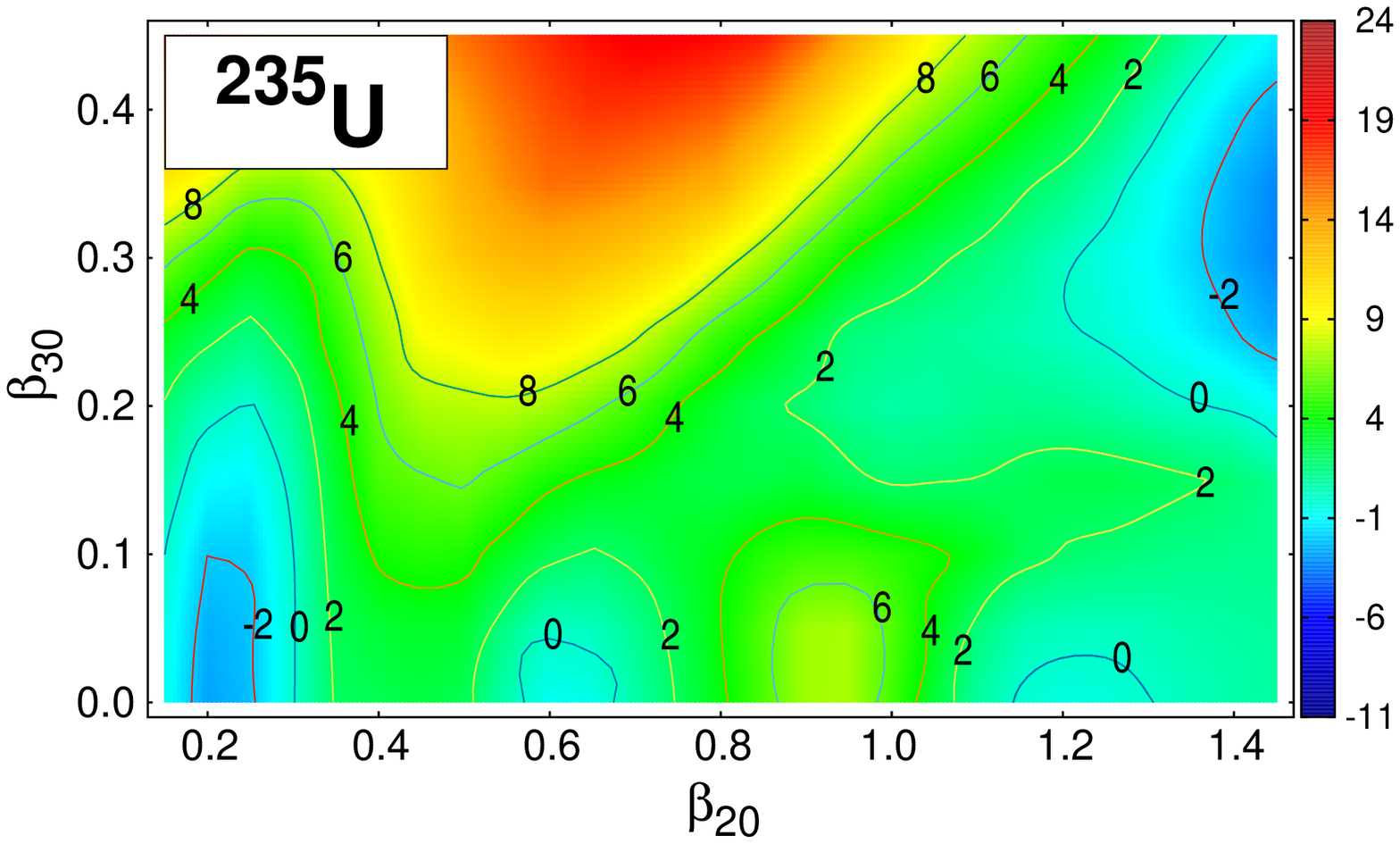}
\includegraphics[scale=0.46]{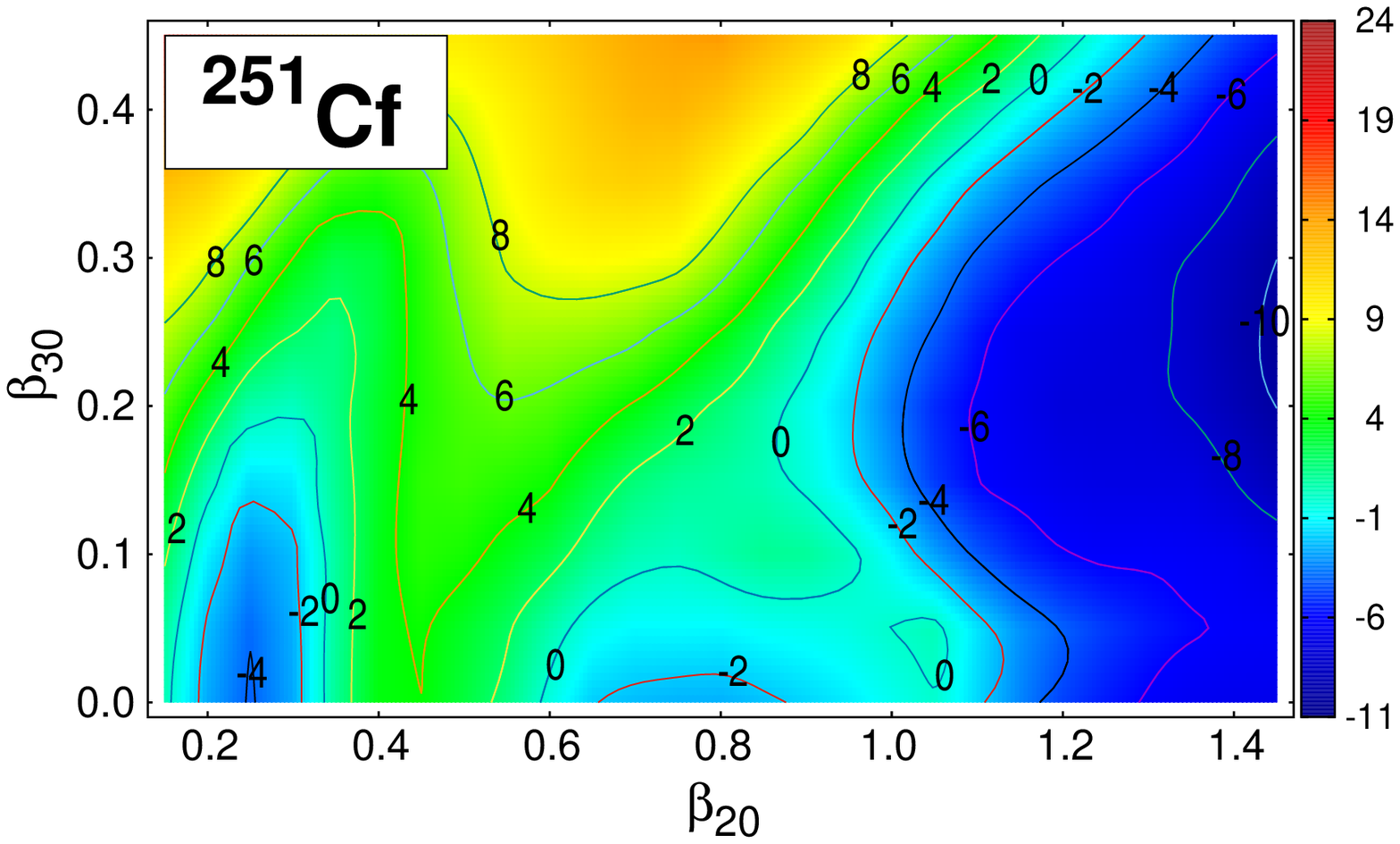}
\caption{The same as in Fig. \ref{fig:Ac_Th} but for $^{235} {\rm U}$ and $^{251} {\rm Cf}$.}
\label{fig:Z_evolution}
\end{figure*}

\subsection{Effect of the pairing-strength increase}  \label{subsection:pairing}

 Here, we address the already mentioned overestimate of the calculated fission
 barriers in odd - $Z$ or/and odd - $N$ systems by a too large effect of
 blocking.
 We stress that we {\it do not} consider an overall (i.e. through all nuclei)
 increase of the pairing strengths. For sure, this would decrease all barriers
 bringing them into a better statistical agreement with the data, but, as
 indicated in subsection III A, at the cost of spoiling the fit to atomic
 masses.

\begin{figure*}[h]
\includegraphics[scale=0.4]{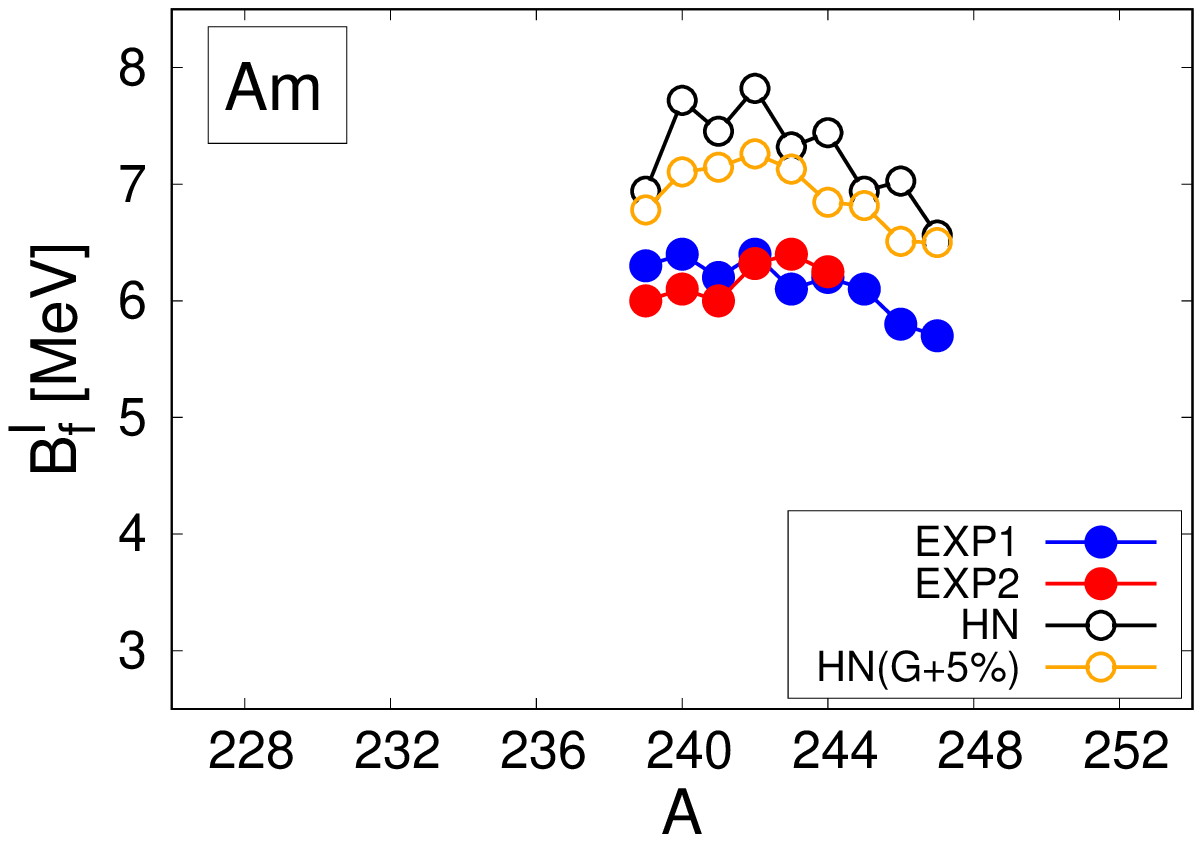}
\includegraphics[scale=0.4]{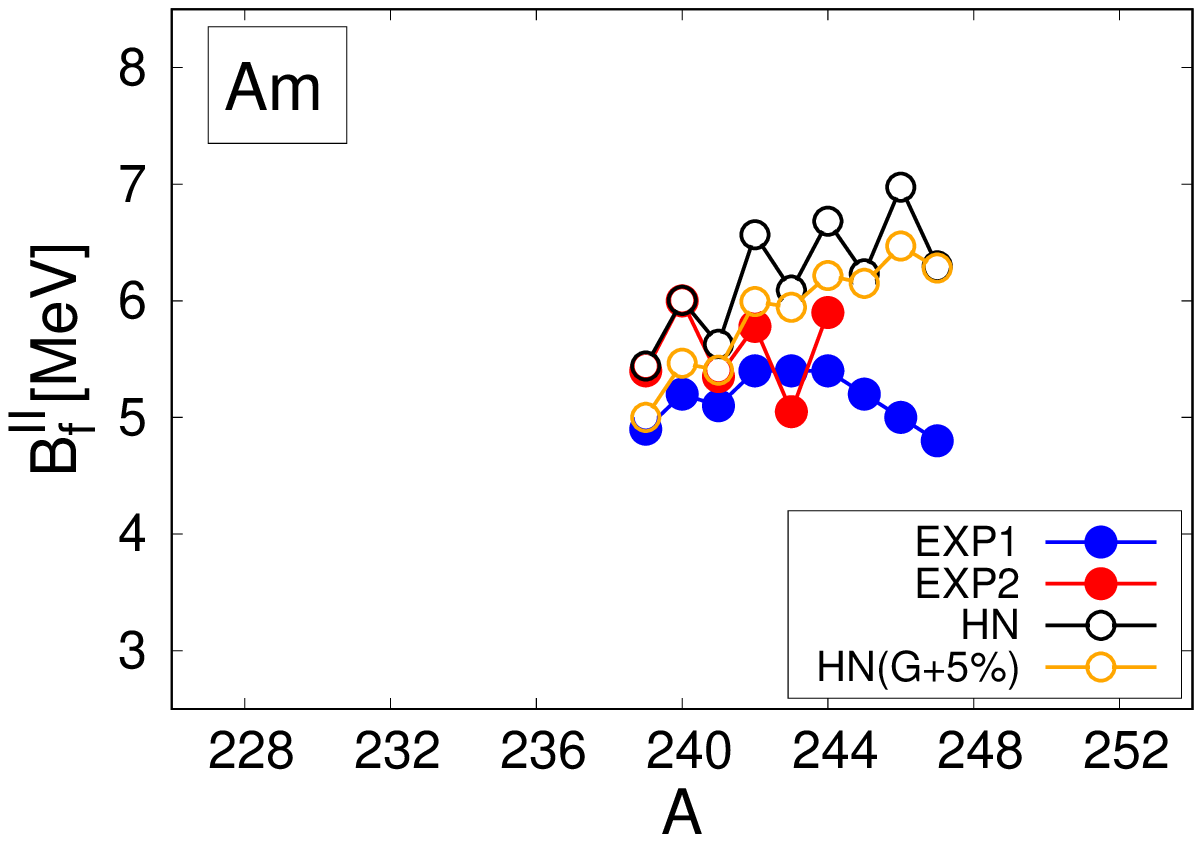}
\caption{First (at left) and second (at right) fission barrier heights
of $^{239-247} {\rm Am}$ isotopes, calculated after increasing
the pairing strength by $5\%$ for odd-particle-number nuclei
(results denoted by orange circles).
Other points are the same as in Figs. \ref{fig:BF1}, \ref{fig:BF2}.}
\label{fig:PAIRING5}
\end{figure*}

 In order to evaluate the effect on the barriers in odd and odd - odd nuclei
 we repeated the whole calculation for Am isotopes with pairing $5\%$ stronger
 for odd protons and odd neutrons.
 The results - inner and outer barrier heights, marked by orange circles,
 are shown in \mbox{Fig. \ref{fig:PAIRING5}}, together with the previous ones
 (black circles) and experimental data.
 As one can see, the calculated barriers are now lowered by up to 0.6 MeV, and
 thus closer to the experimental estimates. This change in Am nuclei
 leads to a decrease in statistical deviations of our barriers from
 the two sets of experimental data given in Tables I and III:
 by $\sim$ 0.05 - 0.07 MeV for the first, and by 0.02 - 0.04 MeV for the second
 barriers.

 A larger increase in pairing strengths for odd-particle number systems would
 lead to the inversion of the odd - even staggering
 in barriers that is not seen in the data, and counter-intuitive in face of
 longer fission half-lives in odd-$A$ nuclei \cite{Hessberger2017}.
 So, the test indicates 0.5 - 0.6 MeV as the maximal possible overestimate of
  barrier heights in odd and odd - odd nuclei introduced by blocking.
 Quite similar conclusion had been obtained earlier in the
 region of superheavy nuclei \cite{Jachimowicz2017_II}.
 It may be mentioned that the applied increase in the pairing strengths
 only moderately increases the discrepancy between calculated and experimental
 g.s. masses - on average by $~ 0.1$, up to 0.3 MeV.

\section{Discussion and Summary}

We have systematically determined inner and outer fission barrier heights
for 75 actinides, within the range from actinium to californium, including
odd-$A$ and odd-odd systems, for which experimental estimates were accessible.
 Obtained barriers are in most cases higher than the experimental estimates.
For odd- and odd-odd nuclei, a (smaller) part of this effect may be a
 consequence of the decrease in the pairing gap due to blocking. Our test
 performed for Am nuclei have shown that blocking can rise barriers by up to
  0.6 MeV, what is consistent with our previous tests and results
 in the region of superheavy nuclei.

 A statistical comparison of our fission barrier heights with available
 experimental estimates gives the average discrepancy and the rms deviation
 not greater than 0.82 MeV and 0.94 MeV, respectively.
This concerns both: first and second fission barriers.
 Determined excitation energies of superdeformed secondary minima reproduce
 quite well the general trends of experimental data. The largest discrepancies
 do not exceed 1.1 MeV.

  The most direct comparison of our results with other calculations is
 possible with \cite{Moller2009} and \cite{Goriely2007}.
  The second fission barriers calculated by P. M\"oller et al. \cite{Moller2009}
  show only a slightly larger statistical deviation from the experimental
 values than ours ($\delta_{rms}=$ 1.07 and 0.90 MeV for the sets I and II,
 respectively); their first barriers are statistically more distant from the
 evaluated data ($\delta_{rms}=$ 1.48 and 1.36 MeV for sets I and II).
 On the other hand, the rms deviations obtained within the HFB Skyrme model
 in \cite{Goriely2007} are astonishingly small, 0.67 and 0.65 MeV, for the
 first and second barriers, respectively. The comparison of results of
 \cite{Goriely2007} and ours is, however, less direct, as
 different nuclei (52 with first barriers, but only part of them actinides,
 and a much smaller number for second barriers - cf Fig. 5 in
 \cite{Goriely2007}) are included. The relatively small
 rms deviation from the experimental data was obtained in \cite{Goriely2007}
 thanks to the subtraction of a purely phenomenological collective correction
 term. Effective mostly at large deformations, it served exclusively to
 correct the Skyrme BSk14 HFB fission barriers, without spoiling the mass
 fit too much.

 Concerning the results for even-even actinides obtained by other authors,
 their agreement with the (smaller number of) data seems to depend on
 corrections applied to the pure mean-field results. Generally, the
 selfconsistent non-relativistic models produce too large barriers if
 one defines them as the energy difference between the saddle and the g.s.
 minimum. They can be brought to a better agreement with experimental
 estimates when additional corrections are applied, like the subtraction
 of the collective rotational energy.
 A very careful analysis of such corrections for SkM* interaction was given
 in \cite{BONNEAU04} and the obtained agreement with experimental barriers illustrated for six actinide nuclei.
 The dependence of results obtained with the
 Gogny force on the assumed corrections, i.e. the way the barrier is
 interpreted, is well documented in
 \cite{Delaroche2006,Robledo2014,Lemaitre2018}. The barriers obtained
 in \cite{Robledo2014} with the D1M interaction for 14 nuclei are
 overestimated by 2-4 MeV. Second barriers obtained in calculations
 with the D1S force \cite{Delaroche2006} are overestimated by 1-2 MeV for $N\geq 144$.
 The first barriers were either too high when calculated in a more standard way, or closer
 to the data when defined, rather arbitrarily, as the energy difference between the $0^{+}$ state
 with the wave function concentrated in the barrier region and the $0^{+}$ ground state.
 In the recent study  \cite{Lemaitre2018}, the first and second barriers in 14 actinide nuclei
 were reproduced with the rms deviation of 0.52 and 0.45 MeV, respectively,
 when the collective energy correction with adiabatic mass parameters was
 applied. It has to be mentioned though
 that the triaxiality was included in a rather crude way in this latter study.
 The relativistic mean-field calculations with the NL3* interaction
  \cite{Abusara2010} reproduced 22 first barriers with the average deviation
 from experimental values of 0.76 MeV. An even better agreement with
 experimental data of calculated 19 first and 15 second barriers was
 obtained in the RMF model with the PC-PK1 interaction in \cite{Zhou2014}.

 One can notice that the overall increase in pairing strengths would bring our
 calculated barriers closer on average (e.g. in the sense of rms deviation) to
 the experimental estimates. However, it would deteriorate the agreement between the calculated and experimental masses
 in actinides.
 Moreover, the statistical improvement would be accompanied by local deteriorations.
 This concerns most of Pa and U isotopes, where calculated
 first fission barriers would become too low vs empirical estimates.
 Already large discrepancies in inner barriers for Th isotopes would increase.

 It should be stressed that some discrepancies seem common to many models.
 It is the case of Th anomaly.
 In calculations, there is a gradual change in widths and heights of inner and
 outer barriers with \mbox{$Z$/$N$}. In Th, inner barriers gain prominence
  with $N$, while in experimental evaluations, high and wide inner barriers
 are assumed in all Th isotopes. As we pointed out, in nearby Ac nuclei,
 calculated PES's are similar to those in Th, while the inner barrier vanishes
 from experimental evaluations. Such an abrupt change in assumptions between
 Ac and Th seems mysterious.

 The other example is an increase with $N$ in the second barriers in Pu and Am,
  resulting from many micro-macro and non-relativistic self-consistent calculations, but not seen in
  data. It seems to point to a more general problem in models or in our
 understanding.

 There is also an intriguing question of third minima, which in our
 calculations, if appear at all, are rather shallow - in most cases
 do not exceed 0.5 - 0.6 MeV in depth. Again, there were experimental
 evaluations claiming much deeper third minima, see e.g.
 \cite{Krasz98,Csige2009}.

 Finally, it seems that while a moderate reduction in deviation of the
 calculated fission barriers from experimental estimates is still possible
 in our and other models, it is not obvious how to achieve it without spoiling
 other observables one would also like to reproduce.


\end{document}